\documentclass[aps,prd,10pt,notitlepage,nofootinbib,superscriptaddress,showkeys,showpacs]{revtex4-1}

\usepackage{amstext,amsmath,amssymb,amsfonts,bbm}
\usepackage[latin1]{inputenc}
\usepackage{epsfig}
\usepackage{dsfont}
\usepackage{hyperref}
\usepackage{amsthm}
\usepackage{subfigure}
\usepackage{color}
\usepackage{multirow}
\usepackage{psfrag}
\usepackage{graphicx}
\usepackage{extarrows}

\theoremstyle{plain}

\theoremstyle{definition}

\newcommand{\beq}{\begin{equation}}
\newcommand{\eeq}{\end{equation}}
\newcommand{\beqa}{\begin{eqnarray}}
\newcommand{\eeqa}{\end{eqnarray}}

\newcommand{\cJ}{{\cal{J}}}

\begin{document}

\title{\Large
Diagrammatics of a colored SYK model and of an SYK-like tensor model, leading and next-to-leading orders}

\author{{\bf Valentin Bonzom}}\email{bonzom@lipn.univ-paris13.fr}
\affiliation{LIPN, UMR CNRS 7030, Institut Galil\'ee, Universit\'e Paris 13, Sorbonne Paris Cit\'e, 99, avenue Jean-Baptiste Cl\'ement, 93430 Villetaneuse, France, EU}

\author{{\bf Luca Lionni}}\email{luca.lionni@th.u-psud.fr}
\affiliation{Laboratoire de Physique Th\'eorique, CNRS UMR 8627, Universit\'e Paris XI, 91405 Orsay Cedex, France, EU}
\affiliation{LIPN, UMR CNRS 7030, Institut Galil\'ee, Universit\'e Paris 13, Sorbonne Paris Cit\'e, 99, avenue Jean-Baptiste Cl\'ement, 93430 Villetaneuse, France, EU}

\author{{\bf Adrian Tanasa}}\email{ntanasa@u-bordeaux.fr}
\affiliation{LaBRI, Universit\'e de Bordeaux, 351 cours de la Lib\'eration, 33405 Talence, France, EU}
\affiliation{H. Hulubei Nat. Inst.  Phys.  Nucl. Engineering, Magurele, Romania, EU}
\affiliation{IUF Paris, France, EU}

\begin{abstract}
The Sachdev-Ye-Kitaev (SYK) model is a model of $q$ interacting fermions. Gross and Rosenhaus have proposed a generalization of the SYK model which involves fermions with different flavors. In terms of Feynman graphs, those flavors are reminiscent of the colors used in random tensor theory. This gives us the opportunity to apply some modern, yet elementary, tools developed in the context of random tensors to one particular instance of such colored SYK models. We illustrate our method by identifying all diagrams which contribute to the leading and next-to-leading orders of the 2-point and 4-point functions in the large $N$ expansion, and argue that our method can be further applied if necessary. In a second part we focus on the recently introduced Gurau-Witten tensor model and also extract the leading and next-to-leading orders of the 2-point and 4-point functions. This analysis turns out to be remarkably more involved than in the colored SYK model. 
\end{abstract}

\maketitle

\section{Introduction}

Sachdev and Ye proposed a toy-model of $N$ spins with Gaussian-random, infinite-range exchange interactions, of order $q$, in $0+1$ dimensions \cite{sy}. This model attracted a certain interest within the condensed matter community. Thus, the 2-point function computation in the large $N$ limit was performed in \cite{PG}.

A simple variant of the Sachdev-Ye model was proposed by Kitaev in a series of seminars \cite{kitaev}. The Sachdev-Ye-Kitaev (SYK) model is a quantum mechanical model of $N$ Majorana fermions living in $0+1$ dimensions with random interactions of order $q$. Kitaev proposed this model as a model of holography.

The SYK model has three remarkable features: it is solvable at strong coupling, 
maximally chaotic and, finally, it presents emergent conformal symmetry. The SYK model is the first model having all these three properties (other known models only have some of these properties, but not all three of them).

This attracted a lot of interest within the high energy physics community. Thus, Maldacena and Stanford studied in detail the two- and four-point functions of the model \cite{MS}, Polchinski and Rosenhaus solved the Schwinger-Dyson equation and computed the spectrum of two-particle states \cite{PR}, Fu {\it et. al.} proposed a supersymmetric version of the model \cite{Fu} and so on.

In \cite{Gross}, Gross and Rosenhaus proposed a generalization of the SYK model. They have included $f$ flavors (which we will most of the time refer to as colors) of fermions, each occupying $N_a$ sites and appearing with a $q_a$ order in the interaction. In the first part of this paper, we will study a colored SYK model which is a particular case of their proposal where each flavor appears only once in the interaction ($q_a = 1$). This colored SYK model (a complex version) was already mentioned in \cite{gurau-ultim}.

Recently, Witten proposed a reformulation of the SYK model using real fermionic tensor fields without quenched disorder \cite{Witten}. This comes from the fact that both the SYK model and random tensor models have the same class of dominant graphs in the large $N$ limit ($N$ being, in the tensor model framework, the size of the tensor). They are known as melonic graphs in the random tensor literature and as (water)melon graphs in the ADS/CFT literature; from a combinatorial point of view, they are series-parallel graphs. 

In \cite{Gurau}, Gurau complemented Witten's results with some modern results of random tensor theory (albeit using a complex version of \cite{Witten}). He gave a classification of the Feynman graphs of the model at all orders in the $1/N$ expansion for the free energy and the 2-point function, based on the Gurau-Schaeffer classification of colored graphs \cite{GS}. Notice that this classification remains somewhat formal, in the sense that it does not give the graphs which contribute at a given order of the $1/N$ expansion. Following the recently introduced terminology, we will refer to the model of \cite{Witten} as the Gurau-Witten model. We will study the Gurau-Witten model in the second part of this paper.

Random tensor models have recently been developed in parallel, in mathematical physics, 
as a natural generalization, in dimensions higher than two, of the celebrated random matrix models. Several classes of random tensor models have been studied in detail in the recent literature: the colored tensor models (see the review paper \cite{GR} or the book \cite{carteRazvan} and references within), the ordinary tensor models which directly generalize one-matrix models (and also known as uncolored model to contrast with the colored model; see for example \cite{uncolored}, \cite{CT} and the review \cite{LargeNSigma} and references within) and the multi-orientable tensor models (see the review paper \cite{mosigma} and references within).

The development of the theory of random tensors is due to combinatorial results on edge-colored Feynman graphs. We have noticed (and so did Gurau in \cite{gurau-ultim} during the completion of this work) that if different colors (or flavors) are added to the fermions which interact in the SYK model (as Gross and Rosenhaus did \cite{Gross}), the Feynman graphs become similar to those of tensor models. \emph{Moreover and crucially, the exponent of $N$ received by each Feynman graph is precisely the same as in (ordinary) tensor models \cite{uncolored}}. This is the starting point of \cite{gurau-ultim} which proves that quenched and annealed disorders are equivalent at leading order in the $1/N$ expansion in this colored SYK model.

In this paper we will study a real version of the colored SYK model. We will use a different approach than in \cite{gurau-ultim}. It is based on very recent combinatorial techniques for edge-colored graphs \cite{blr} but we will only need some elementary version of \cite{blr}. Those techniques have been developed in order to go (successfully to some extent \cite{LargeNSigma}) beyond the melonic phase in ordinary tensor models. We will here explain in detail how to extract the leading order (LO), the next-to-leading order (NLO) of the 2-point and 4-point functions using the most simple version of those techniques. The LO reproduces trivially melons and chains (known as ladders in \cite{MS}), and give new graphs at NLO. We even included some details on the next-to-next-to-leading order (NNLO) of the 2-point function to show that our method is fairly straightforward to apply. 

The Gurau-Witten model has attracted a certain interest, coming from the fact that this model, while having a similar behavior to the SYK model at large $N$, does not require averaging over quenched disorder. Thus, Klebanov and Tarnolopsky proposed in \cite{KT} several SYK-like tensor models based on the (uncolored) 1-tensor model of \cite{CT} and on the multi-orientable tensor model \cite{mosigma}. Peng {\it et. al.} proposed a supersymmetric version of a Gurau-Witten model \cite{brown}. Numerical analysis was performed in \cite{KSS} on the simplest instance of the Gurau-Witten, to compare the spectrum of its Hamiltonian to the SYK model. Closely related to the tensor approach is the consideration of a large number $D$ of $N\times N$ matrices, where the large $D$ limit is taken \cite{ulb}. This idea had been originally brought forward in \cite{bc} to think of tensor models as multi-matrix models with a special, large $D$, limit (the symmetry group is actually different in \cite{ulb} and \cite{bc}).

Let us insist that the colored SYK model and the Gurau-Witten model are different although they coincide at leading order in the large $N$ limit. This is easily seen by looking at their perturbative expansions. At fixed couplings in the colored SYK model, the Feynman graphs are the same as those of the Gurau-Witten model. However, upon averaging over the disorder, every such graph has to be supplemented with a sum over additional Wick pairings, which do not affect the Feynman amplitudes but determine the weights in the $1/N$ expansion. This produces copies of it which typically receive different $1/N$ weights. This means that a graph of the Gurau-Witten model appears in the colored SYK model at various orders of the $1/N$ expansion. In addition, both models assign $N$-dependent weights to graphs which count different types of cycles.

\medskip

The paper is organized as follows. In the next section (section \ref{sec:GeneralizedSYK}) we recall the Gross-Rosenhaus generalization of the SYK model and specify the version we will study. In section \ref{sec:SYK}, we show what the LO and NLO vacuum, two- and four-point diagrams are, using a simple method alternative to \cite{gurau-ultim}. The same analysis is performed in section \ref{sec:GW} for the real Gurau-Witten tensor model. Some tables which summarize our results are given in section \ref{sec:Summary}. We offer some concluding remarks in section \ref{sec:Conclusion}, comparing the diagrammatics of the two models we studied at NLO in particular.

Figures and graphs are represented with $q=4$ throughout the text, i.e. Feynman graphs with $q=4$ fermions interacting at each vertex. We only discuss the diagrammatics of the $1/N$ expansions and leave out completely the Feynman amplitudes.

\section{Generalisations of SYK -- the colored SYK models} \label{sec:GeneralizedSYK}

The SYK model contains $N$ real fermions $\psi_i$ ($i=1,\ldots, N$), with $q-$fold random coupling, $q$ being here an even integer.
The action writes:
\beqa
\label{act:syk}
S_{\mathrm{SYK}}=\int d\tau \left( 
\frac 12 \sum_{i=1}^N \psi_i \frac{d} {dt} \psi_i
- \frac{i^{q/2}}{q!}
\sum_{i_1,\ldots, i_q=1}^N
j_{i_1\ldots i_q}\psi_{i_1}\ldots \psi_{i_q}
\right)
\eeqa
Note that the most widely studied version of the SYK model has real fermions, as above. This was done in \cite{kitaev} and subsequent papers.

\begin{figure}
\begin{center}
\includegraphics[scale=0.9]{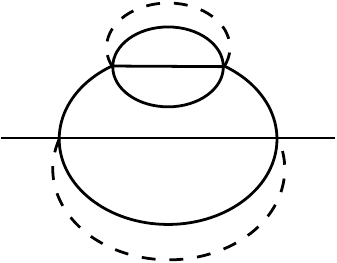}\hspace{2cm} \includegraphics[scale=0.9]{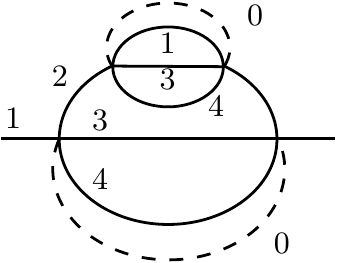}
\caption{\label{melonSYK}Melonic graphs of the SYK and colored SYK models}
\end{center}
\end{figure}

As an example of Feynman graphs obtained by perturbative expansion, we show on the left of Fig. \ref{melonSYK} a melonic graph, which is a dominant graph in the large $N$ expansion of the SYK model. As usually, the dashed line represents the disorder.

The SYK generalization we study contains $q$ flavors of fermions. Each fermion of a given flavor appears exactly once in the interaction and the Lagrangian couples $q$ fermions together. The action is:
\begin{equation}
\label{act:noi}
S=\int d\tau \left( 
\frac 12 \sum_{f=1}^q\sum_{i=1}^N \psi_i^f \frac{d} {dt} \psi_i^f
- \frac{i^{q/2}}{q!}\sum_{i_1,\ldots, i_q=1}^N
j_{i_1\ldots i_q}\psi_{i_1}^{1}\ldots \psi_{i_q}^{q},
\right)
\end{equation}
Note that we use superscripts to denote the flavor. Moreover, to easy notations, we now work with $q\cdot N$ fermions - we have $N$ fermions of a given flavor.

\medskip

The SYK generalization introduced above is a particular case of the Gross-Rosenhaus generalization  \cite{Gross}. Indeed, Gross and Rosenhaus took a number $f$ of flavors, with $N_a$ fermions of flavor $a$, each appearing $q_a$ times in the interaction, such that $N=\sum_{a=1}^f N_a$ and $q=\sum_{a=1}^f q_a$. In our case, the number $f$ of flavors is equal to $q$, $q_a=1$ and $N_a=N$ (recall that we have now a total of $q\cdot N$ fermions).

\medskip

Thus, the Feynman graphs obtained through perturbative expansion of the action \eqref{act:noi} are edge-colored graphs where the colors are the flavors. At each vertex, each of the $q$ fermionic fields which interact has one of the $q$ flavors, and each flavor is present exactly once.

An example of such a Feynman graph is given on the right of Fig. \ref{melonSYK}. 
Once again, we have represented the disorder by a dashed line. This line can be considered to have the fictitious flavor $0$.

\medskip

In this paper we will focus on the real model \eqref{act:noi}. There is also a complex version \cite{gurau-ultim}. This latter version can be easily obtained by considering the propagation from $\psi$ to $\bar \psi$ and by considering the interacting term in \eqref{act:noi} as well as its complex conjugate:
\begin{equation}
\label{act:noir}
\int d\tau \left( 
\frac 12 \sum_{f=1}^q\sum_{i=1}^N \bar \psi_i^f \frac{d} {dt} \psi_i^f
- \frac{i^{q/2}}{q!}\sum_{i_1,\ldots, i_q=1}^N
j_{i_1\ldots i_q}\psi_{i_1}^{1}\ldots \psi_{i_q}^{q}
- \frac{(-i)^{q/2}}{q!}\sum_{i_1,\ldots, i_q=1}^N
\bar j_{i_1\ldots i_q}\bar \psi_{i_1}^{1}\ldots \bar\psi_{i_q}^{q},
\right)
\end{equation}
The Feynman graphs obtained through perturbative expansion of the complex action have the same structure as the one explained above for the real model \eqref{act:noi}. However, in the complex case, one has two types of vertices, which we can refer to as white and black, as it is done in the tensor model literature (see again the book \cite{carteRazvan} and references within). Each line connects a white to a black vertex. The Feynman graphs of \eqref{act:noir} are thus the subset of the Feynman graphs of \eqref{act:noi} which are bipartite. This is a feature which  simplifies the diagrammatic analysis of the model. In this paper however, we will have to deal with typically non-bipartite graphs.

\section{Diagrammatics of the colored SYK model} \label{sec:SYK}

At fixed couplings $j_{i_1 \dotsc i_q}$, the Feynman graphs are $q$-regular edge-colored graphs: graphs with a color from $\{1, \dotsc, q\}$ on each edge and such that all colors are incident exactly once one each vertex.

In order to study the $1/N$ expansion, one must average over the disorder with the covariance
\begin{equation}
\langle j_{i_1 \dotsc i_q} j_{l_1 \dotsc l_q} \rangle \sim \frac1{N^{q-1}} \prod_{k=1}^q \delta_{i_k, l_k}
\end{equation}
and similarly for $\langle j_{i_1 \dotsc i_q} \bar{j}_{l_1 \dotsc l_q} \rangle$ in the complex case. Each graph is thus turned into a sum over Wick pairings which can be represented with edges carrying a new color, say the color $0$.

This adds to the fermionic Feynman graphs an additional set of edges with color 0. An edge of color 0 must join two vertices. For each color $i\in\{1, \dotsc, q\}$, a graph has cycles (i.e. closed paths) which alternate the colors 0 and $i$. They are called \emph{faces of colors $0i$}. This terminology is an extension of matrix models where those cycles are faces of ribbon graphs.

We denote $F_{0i}(G)$ the number of faces of colors $0i$ for $i=1, \dotsc, q$ of a graph $G$, 
\begin{equation}
F_0(G) = \sum_{i=1}^q F_{0i}(G)
\end{equation}
the total number of faces which have the color 0, and $E_0(G)$ its number of edges of color 0 (obviously half the number of vertices of $G$). It is easy to see that $G$ has a free sum for each face of colors $0i$ which sums up to $N$. It thus receives a total weight
\begin{equation}
\label{SYKdeg}
w_N(G) = N^{\chi_0(G)} \qquad \text{with $\chi_0(G)= F_0(G) - (q-1)E_0(G)$}.
\end{equation}
Gurau's theorem on the $1/N$ expansion of tensor models (see, for example, the version presented in \cite{uncolored}) ensures that this is bounded,
\begin{equation}
\chi_0(G) = F_0(G) - (q-1)E_0(G) \leq \begin{cases} 1 & \text{if $G$ is a vacuum graph,}\\
0 & \text{if $G$ is a 2-point graph.} \end{cases}
\end{equation}
The case of 4-point graphs will be discussed later. 
In the language of \cite{uncolored}, the graphs of the colored SYK models have a single bubble, i.e. a single connected component after removing the edges of color 0, since this bubble is the underlying, connected fermionic graph at fixed couplings.

\subsection{LO, NLO of vacuum and 2-point graphs} \label{sec:2Pt}

Notice that all 2-point graphs are obtained by cutting an edge $e$ of color $i\in\{1, \dotsc, q\}$ in a vacuum graph $G$. Since there is a single face, with colors $0i$, which goes through $e$ in $G$, cutting it decreases the exponent of $N$ by one exactly.

To study $\chi_0(G)$, we perform in $G$ the contraction of the edges of color 0 to get the graph $G_{/0}$,
\begin{equation}
\label{Contraction}
\begin{array}{c} \includegraphics[scale=.6]{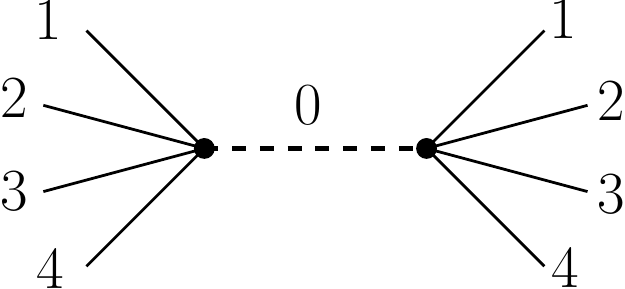} \end{array} \qquad \underset{/0}{\to} \qquad \begin{array}{c} \includegraphics[scale=.6]{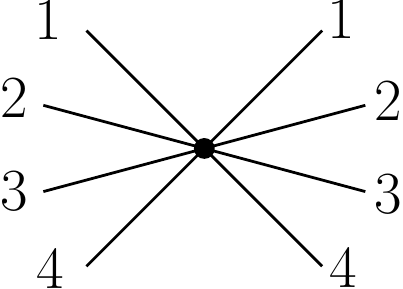} \end{array}
\end{equation}
This means that two vertices of $G$ connected by an edge of color 0 become a single vertex in $G_{/0}$. The map $G\mapsto G_{/0}$ is not one-to-one because of this. Nevertheless, in the complex case, where $G$ is bipartite, it can be made one-to-one by orienting the edges from, say, $\psi^i$ to $\bar{\psi}^i$, i.e. from white to black vertices. Then the edges of $G_{/0}$ are oriented and this is sufficient to reconstruct $G$. In the real case, $G$ is not always bipartite and there are typically several graphs $G$ for the same $G_{/0}$.

\emph{The main property of $G_{/0}$ is that all $q$ colors are incident exactly twice on each vertex}. Therefore, the edges of color $i$ form a disjoint set of cycles (we recall that a cycle is a closed path which visits its vertices only once). Let $\ell_i(G_{/0})$ be the number of cycles of edges of color $i$. From the construction of $G_{/0}$, its cycles of color $i$ are the faces of colors $0i$ of $G$,
\begin{equation}
F_{0i}(G) = \ell_i(G_{/0}).
\end{equation}
Let us introduce $L(G_{/0})$ the cyclomatic number of $G_{/0}$, i.e. its number of independent cycles, or first Betti number. As is well known, it is the number of edges of $G_{/0}$ minus its number of vertices plus one. The number of edges of $G_{/0}$ is the number of edges of $G$ with colors in $\{1, \dotsc, q\}$, thus $q E_0(G)$. The number of vertices of $G_{/0}$ simply is $E_0(G)$, so that
\begin{equation}
L(G_{/0}) = (q-1) E_0(G) + 1.
\end{equation}
This shows that 
\begin{equation}
\chi_0(G) = \sum_{i=1}^d \ell_i(G_{/0}) - L(G_{/0}) + 1
\end{equation}
which has a simple graphical interpretation: it is minus the number of multicolored cycles. Indeed, a cycle can be single-colored or multicolored. The former are counted by $\sum_{i=1}^q \ell_i(G_{/0})$ while $L(G_{/0})$ counts the total number of cycles. Therefore their difference leaves precisely the number of cycles $\ell_m(G_{/0})$ which are multi-colored, up to a sign,
\begin{equation} \label{DegreeCycles}
\chi_0(G) = - \ell_m(G_{/0}) + 1.
\end{equation}
The classification of graphs $G$ with respect to $\chi_0(G)$ is therefore obtained from $\ell_m(G_{/0})$.

\subsubsection{Leading order}

The large $N$ limit consists in graphs such that $\ell_m(G_{/0}) = 0$, i.e. $G_{/0}$ has no multicolored cycles. It means that it is made of single-colored cycles which are glued without forming additional cycles. The corresponding graphs $G$ are easily seen to be melonic. Indeed, one starts from $G_{/0}$ being a simple single-colored cycle of color $i\in\{1, \dotsc, q\}$ with loops of all other colors on its vertices. Then each vertex of $G_{/0}$ is replaced with a pair of vertices and each loop becomes an edge between them. The color 0 from the average over disorder is added between the vertices of each pair too. One gets a melonic cycle as follows,
\begin{equation}
G_{/0} = \begin{array}{c} \includegraphics[scale=.6]{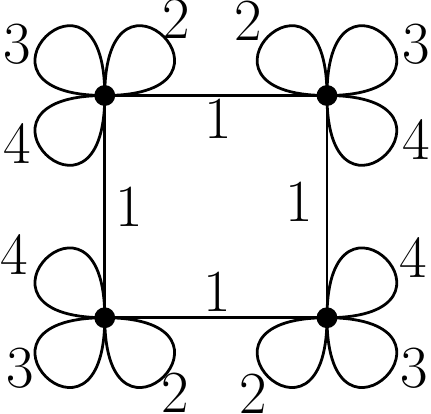} \end{array} \qquad \Rightarrow \qquad G = \begin{array}{c} \includegraphics[scale=.6]{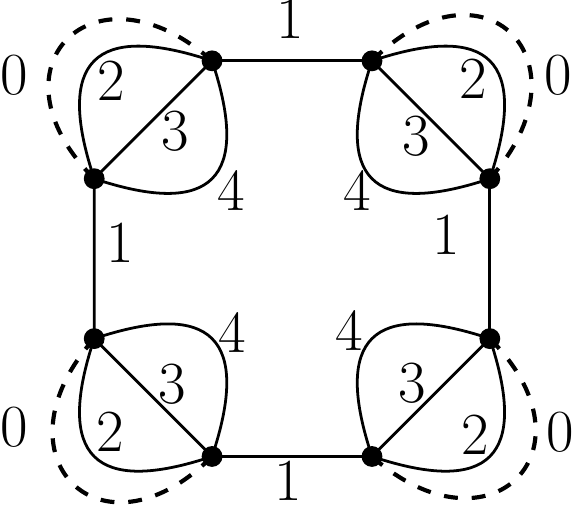} \end{array}.
\end{equation}
More general $G_{/0}$ are obtained by cutting a loop, say of color 2, and replacing it with a cycle and loops attached to its vertices. This corresponds to cutting an edge of color 2 in $G$ and gluing another melonic cycle. This recursive process generates all the graphs corresponding to the large $N$ limit.

The large $N$ 2-point function is simply obtained by cutting an edge of color $i\in\{1, \dotsc, q\}$. From the above recursive process, one finds the following description of the large $N$, fully dressed propagator
\begin{equation} \label{Melonic2Pt}
\begin{array}{c} \includegraphics[scale=.6]{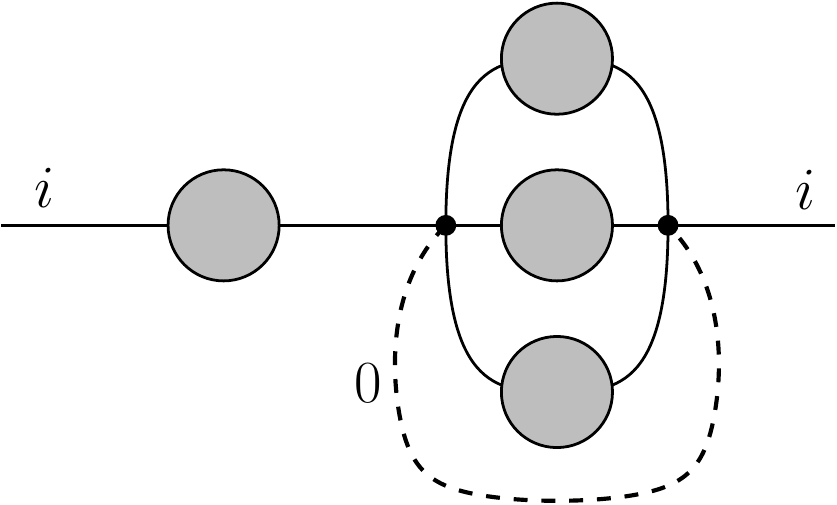} \end{array}
\end{equation}
where each gray blob reproduces the same structure.

\subsubsection{Next-to-leading order and chains}

It is easy to check that replacing an edge in $G$ with any LO 2-point function of the form \eqref{Melonic2Pt} does not change $\chi_0(G)$. Therefore, \emph{all solid edges in the remaining of the article are large $N$, fully dressed propagators}.

2-point functions in the representation as $G_{/0}$ are simply obtained by contracting all edges of color 0 of 2-point graphs $G$. Therefore, solid edges in $G_{/0}$ will also represent fully dressed propagators from now on.

At NLO, one finds graphs such that $\ell_m(G_{/0}) = 1$, i.e. $G_{/0}$ has a single multicolored cycle. Compared to the large $N$ limit, this means that one obtains $G_{/0}$ by gluing single-colored cycles (with loops attached to their vertices) so as to form a single multicolored cycle.

Considering that solid edges are fully dressed 2-point functions, the NLO graphs $G_{/0}$ are completely characterized by the length $n$ of the multicolored cycle with colors $i_1, i_2, \dotsc, i_n$. For instance at length $n=6$:
\begin{equation}
G_{/0}^{\text{NLO}} = \begin{array}{c} \includegraphics[scale=.6]{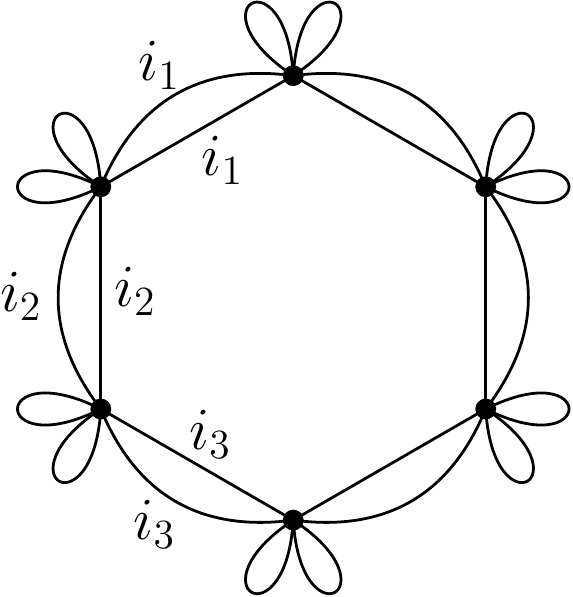} \end{array}
\end{equation}

To find the corresponding graphs $G$, one splits each vertex of $G_{/0}$ into two vertices connected by an edge of color 0 and so that each color is incident exactly once on each vertex. There are several ways to connect the edges of color $i_j$ and $i_{j+1}$ to a pair of vertex. Overall, this leads to the two following families of graphs,
\begin{equation} \label{VacuumNLO}
G^{\text{NLO}} = \begin{array}{c} \includegraphics[scale=.5]{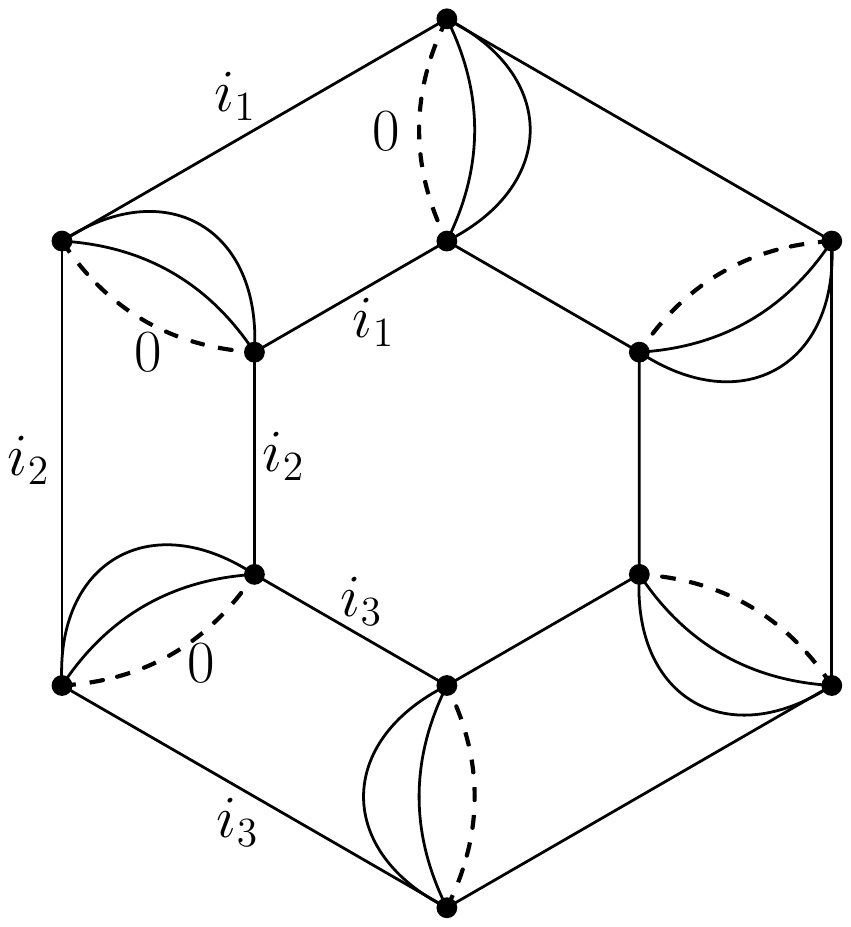} \end{array} \qquad \text{and} \qquad \tilde{G}^{\text{NLO}} = \begin{array}{c} \includegraphics[scale=.5]{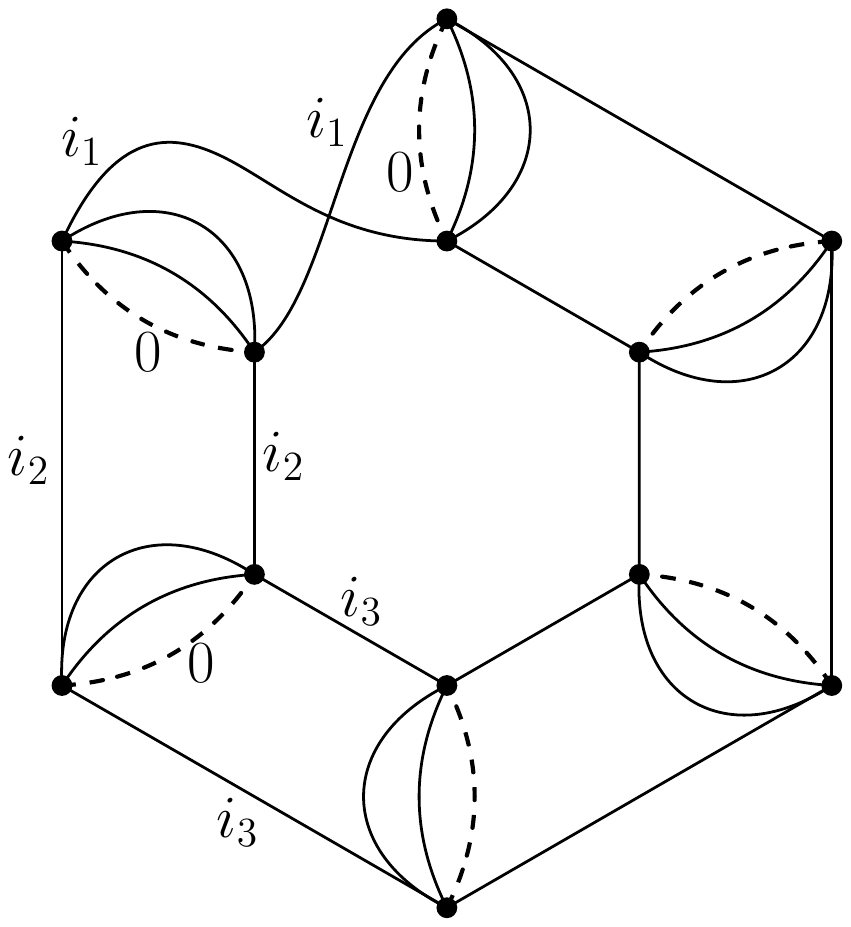} \end{array}
\end{equation}
Notice that $G^{\text{NLO}}$ is bipartite (recall that the melonic 2-point functions are) while $\tilde{G}^{\text{NLO}}$ is not.
The graph $\tilde{G}^{\text{NLO}}$ is obtained from $G^{\text{NLO}}$ by crossing two edges, say with colors $i_1$. Adding more crossings is always equivalent to $G^{\text{NLO}}$ (for an even number of crossings) or $\tilde{G}^{\text{NLO}}$ (for an odd number of crossings).

To remember that the graphs above can have arbitrary lengths, and also to offer a convenient representation of NLO 2-point functions (to come below), we introduce \emph{chains} which are 4-point graphs,
\begin{equation} \label{SYKChain}
\begin{array}{c} \includegraphics[scale=.6]{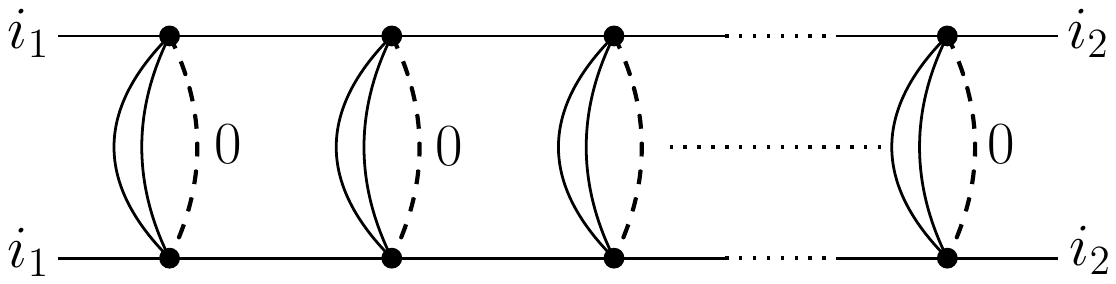} \end{array}
\end{equation}
A combinatorial detail of importance is that a chain can have down to two vertices only, and has at least two vertices unless stated otherwise. 
We will represent arbitrary choices of chains as boxes,
\begin{equation} \label{SYKChainVertex}
\begin{array}{c} \includegraphics[scale=.6]{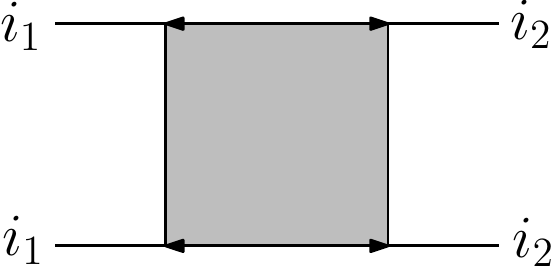} \end{array}
\end{equation}
where the arrows indicate the direction of the chain (the box by itself being symmetric).

This enables to represent the two families of NLO vacuum graphs as
\begin{equation}
\label{Vacuum_NLO_Schemes}
G^{\text{NLO}} = \begin{array}{c} \includegraphics[scale=.5]{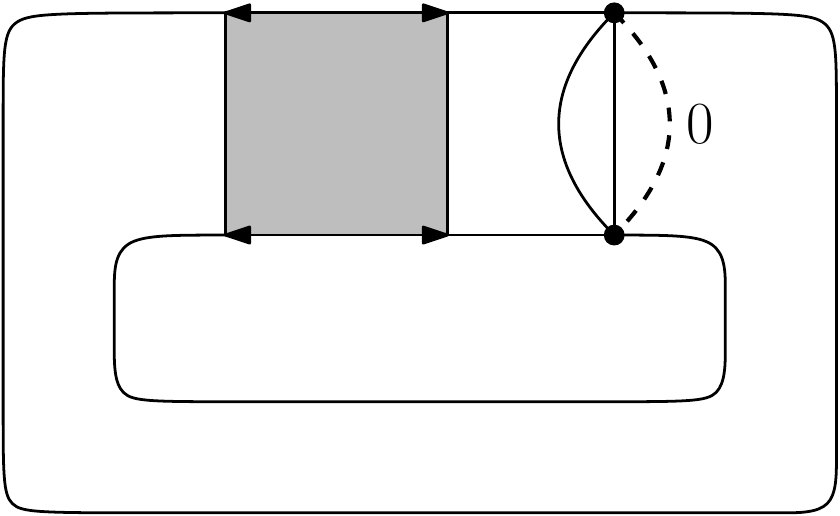} \end{array} \qquad \text{and} \qquad \tilde{G}^{\text{NLO}} = \begin{array}{c} \includegraphics[scale=.5]{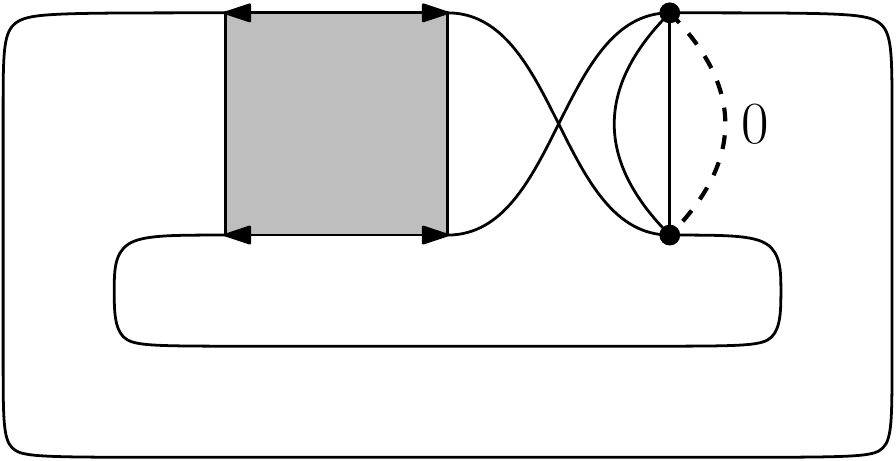} \end{array}
\end{equation}

To get the 2-point functions at NLO from vacuum graphs, it is sufficient to cut an edge of a given color $i\in\{1, \dotsc, q\}$ in a NLO vacuum graph. However, we have to remember that we have used dressed propagators in \eqref{VacuumNLO}. For instance, $G^{\text{NLO}}$ really is
\begin{equation} \label{NLOGraph}
\begin{array}{c} \includegraphics[scale=.6]{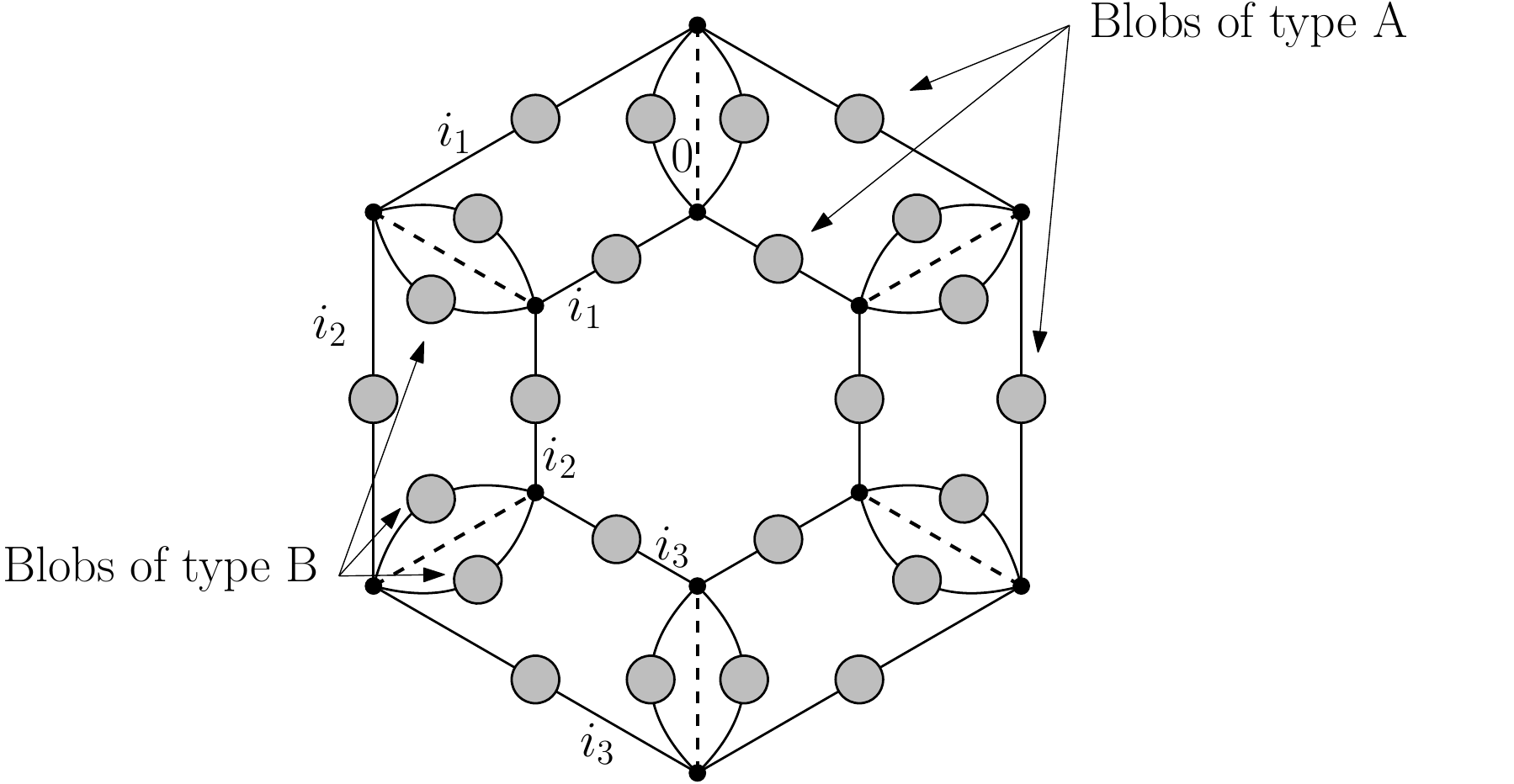} \end{array}
\end{equation}
where the gray blobs represent arbitrary, LO 2-point functions. One might (not necessarily but typically) cut an edge which is contained in a gray blob of \eqref{NLOGraph}. There are two cases to distinguish depending on where an edge is cut in \eqref{NLOGraph}, because there are two types of blobs in \eqref{NLOGraph}.
\begin{itemize}
\item Blobs of type A are inserted on the $2n$ edges of colors $i_1, \dotsc, i_n$ which are characterized as follows: such an edge connects two vertices which are not incident to the same edge of color 0.
\item Blobs of type B are the others: they are inserted on the edges whose end-points are incident to the same edge of color 0.
\end{itemize}

If the cut edge is chosen within a blob of type A, then there are two types of NLO 2-point graphs:
\begin{equation} \label{NLO2Pt1}
G_2^{\text{NLO}(1)} = \begin{array}{c} \includegraphics[scale=.45]{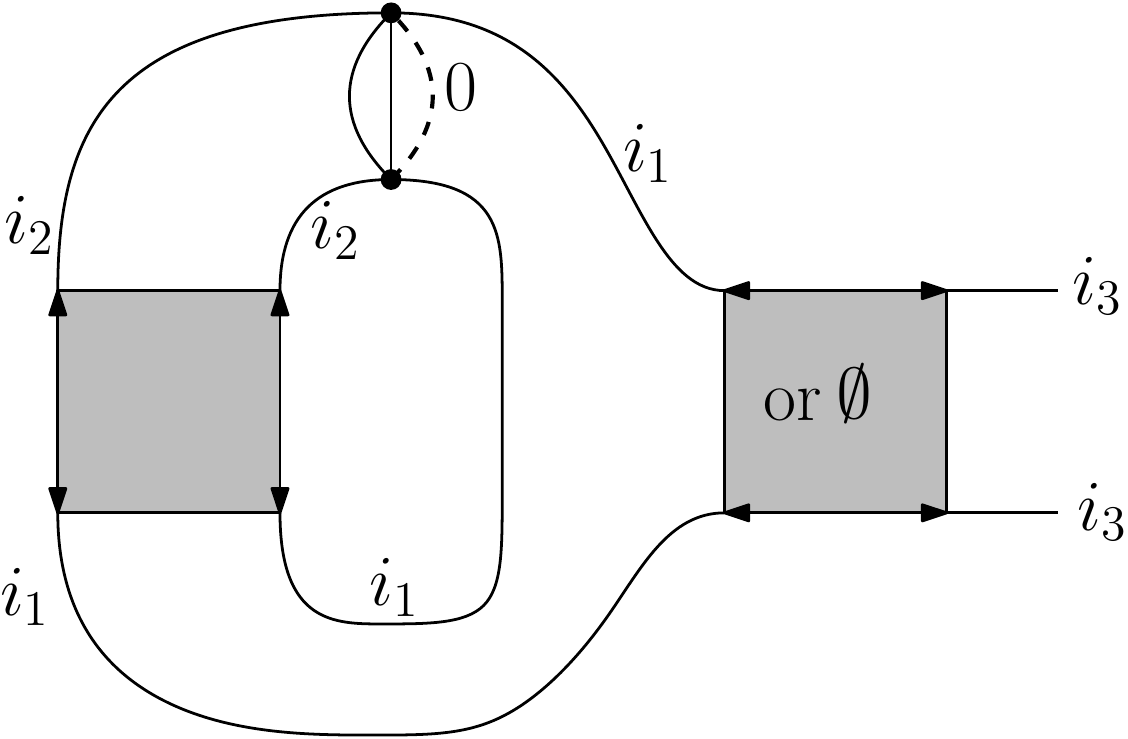} \end{array},
\tilde{G}_2^{\text{NLO}(1)} = \begin{array}{c} \includegraphics[scale=.45]{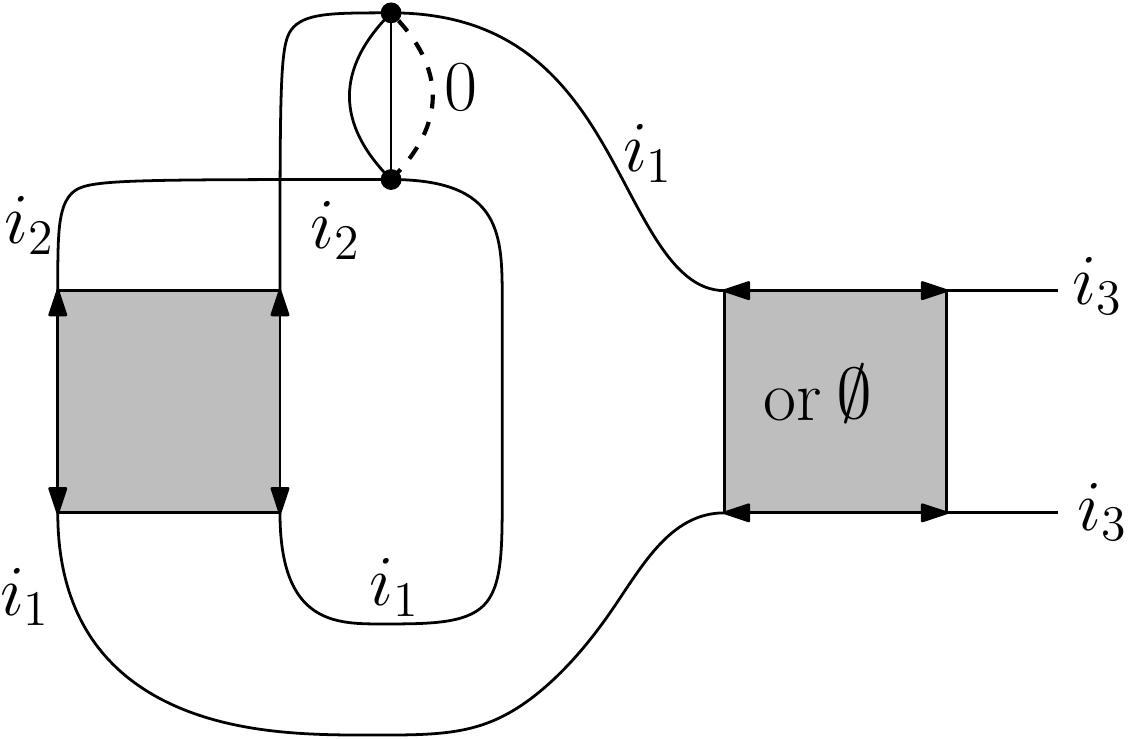} \end{array}
\end{equation}
Only $G_2^{\text{NLO}(1)}$ is bipartite (and would thus contribute in a complex model).

If the cut edge is within a blob of type B, then the NLO 2-point contributions are
\begin{equation} \label{NLO2Pt2}
G_2^{\text{NLO}(2)} = \begin{array}{c} \includegraphics[scale=.45]{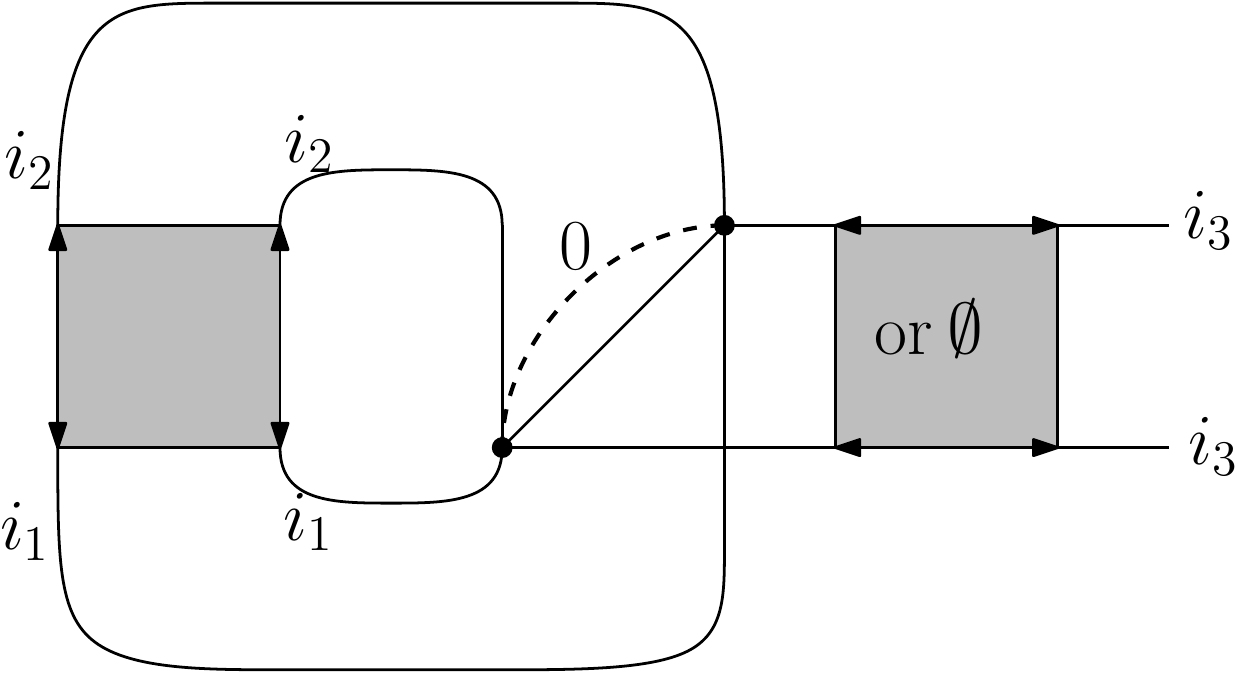} \end{array},
\tilde{G}_2^{\text{NLO}(2)} = \begin{array}{c} \includegraphics[scale=.45]{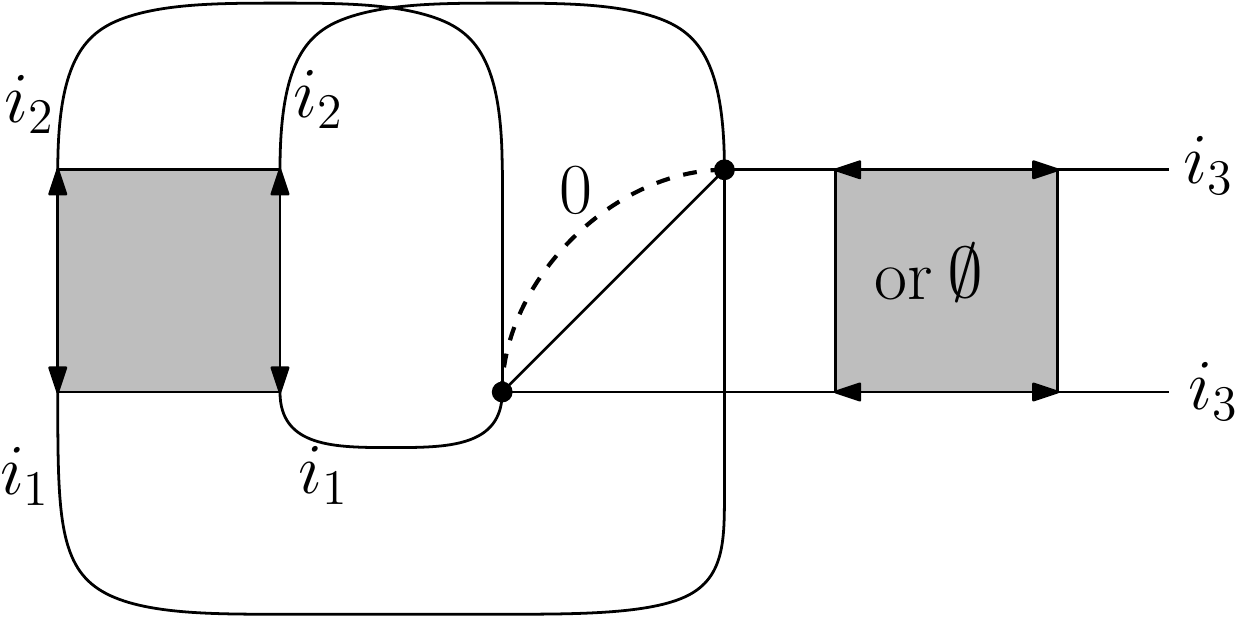} \end{array}
\end{equation}
Again, only $G_2^{\text{NLO}(2)}$ is bipartite.

\subsubsection{Following orders of the partition function}

The method we have used to identify LO and NLO contributions to the free energy and 2-point function can in principle be applied at any order. However, the number of diagrams grows importantly and the description becomes tedious. Here we therefore only give the diagrams which contribute to the NNLO of the partition function.

Graphs contributing to the NNLO are such that the corresponding $G_{/0}$ have exactly two independent multicolored cycles,
\begin{equation}
\ell_m(G^{\text{NNLO}}_{/0}) = 2.
\end{equation}
A reasoning similar to the NLO case of Section \ref{sec:2Pt} leads to families of graphs such as the following ones
\begin{equation} \label{VacuumNNLO}
\begin{array}{c} \includegraphics[scale=.4]{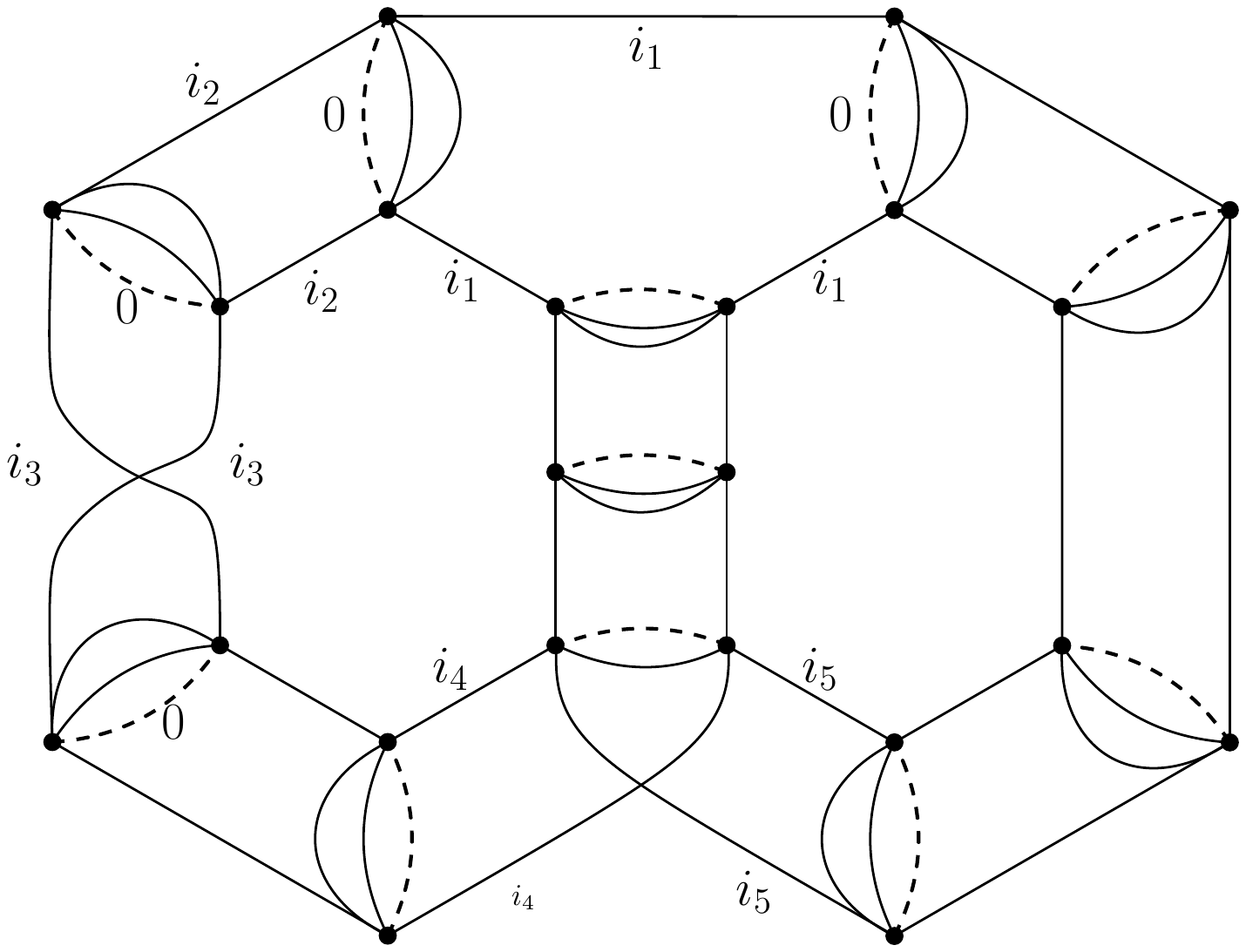} \end{array} \quad \text{and} \quad 
\begin{array}{c} \includegraphics[scale=.4]{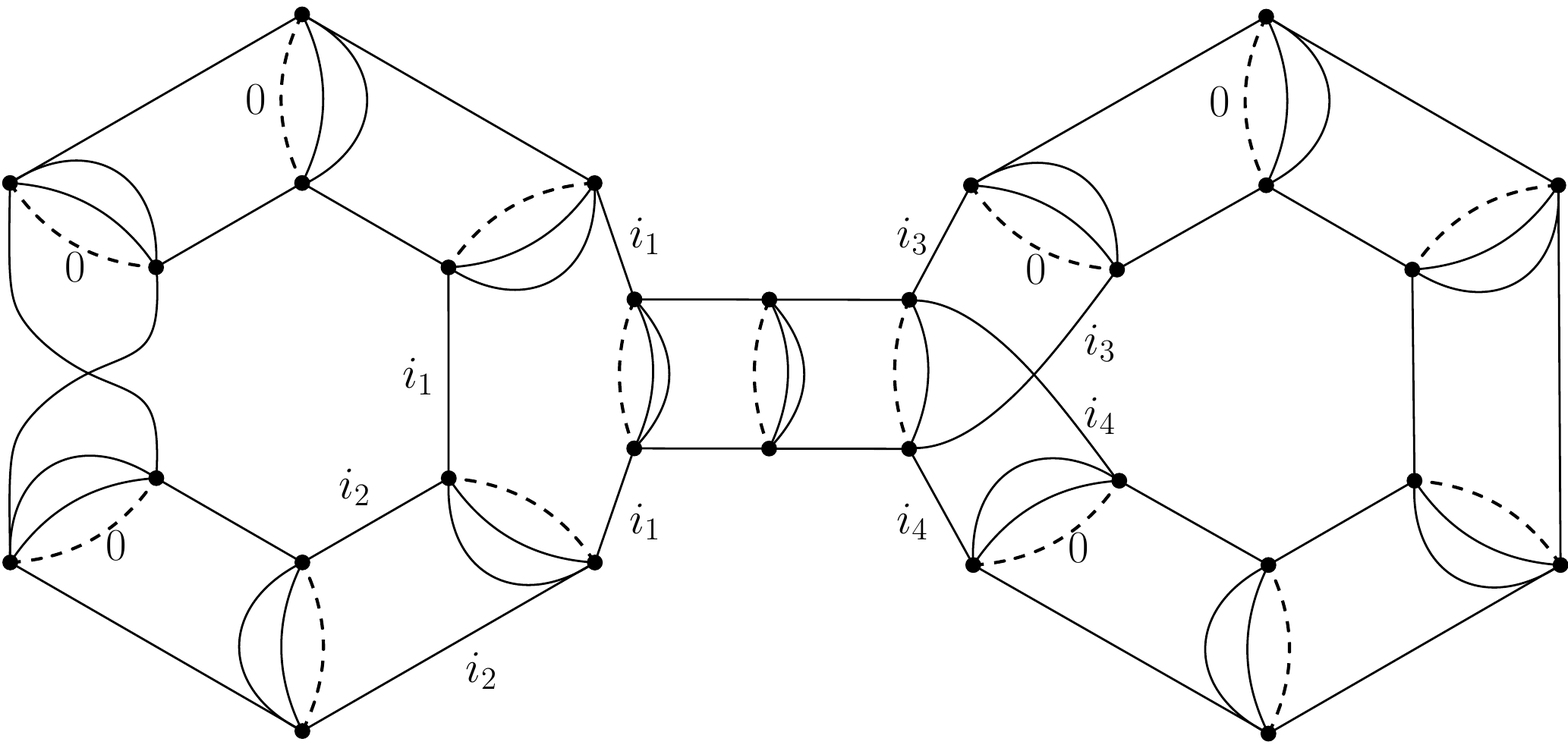} \end{array}
\end{equation}
A collection of diagrams is pictured below. To obtain all such graphs, one has to consider one crossing or no crossing in every loop in every possible way.
\begin{equation}
\label{SYK_Vacuum_NNLO1}
\begin{array}{c} \includegraphics[scale=.35]{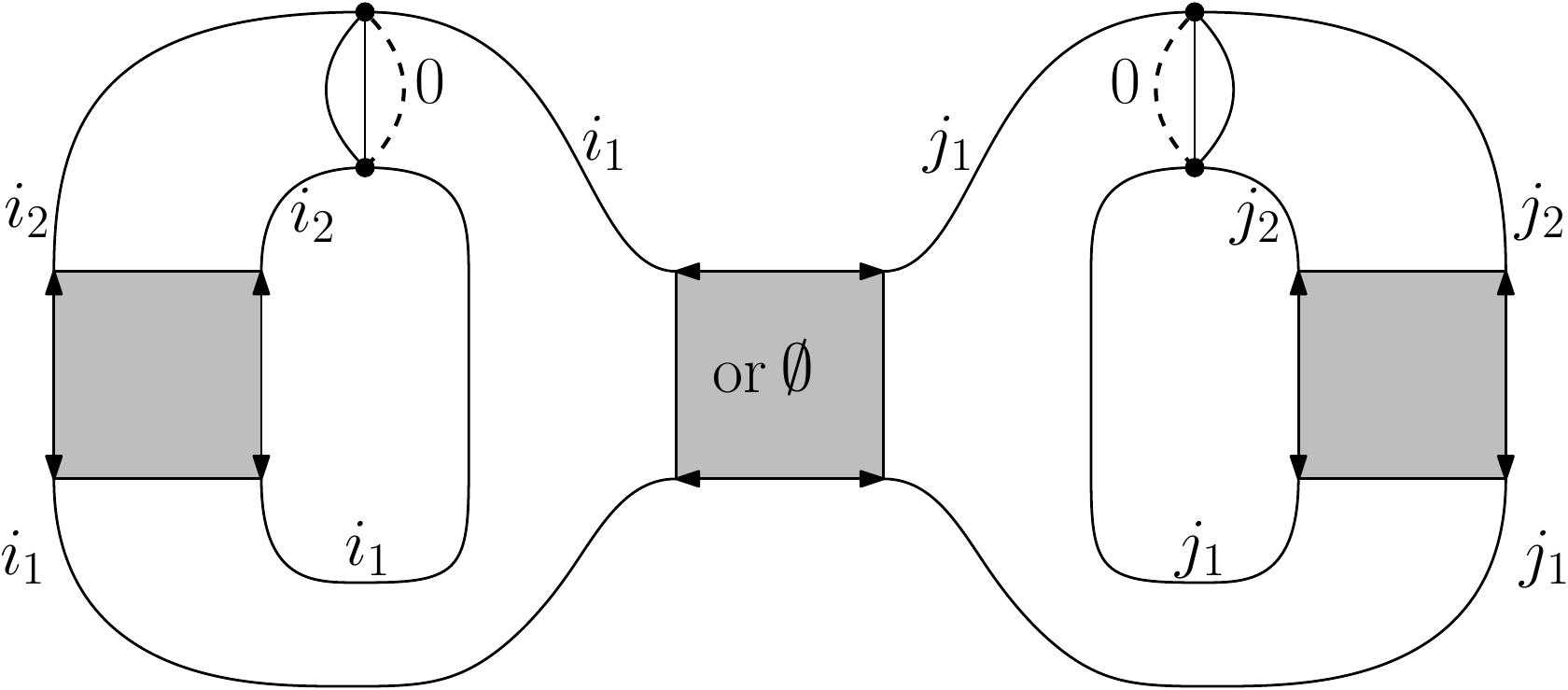} \end{array} \hspace{2.2cm}  
\begin{array}{c}
\includegraphics[scale=.35]{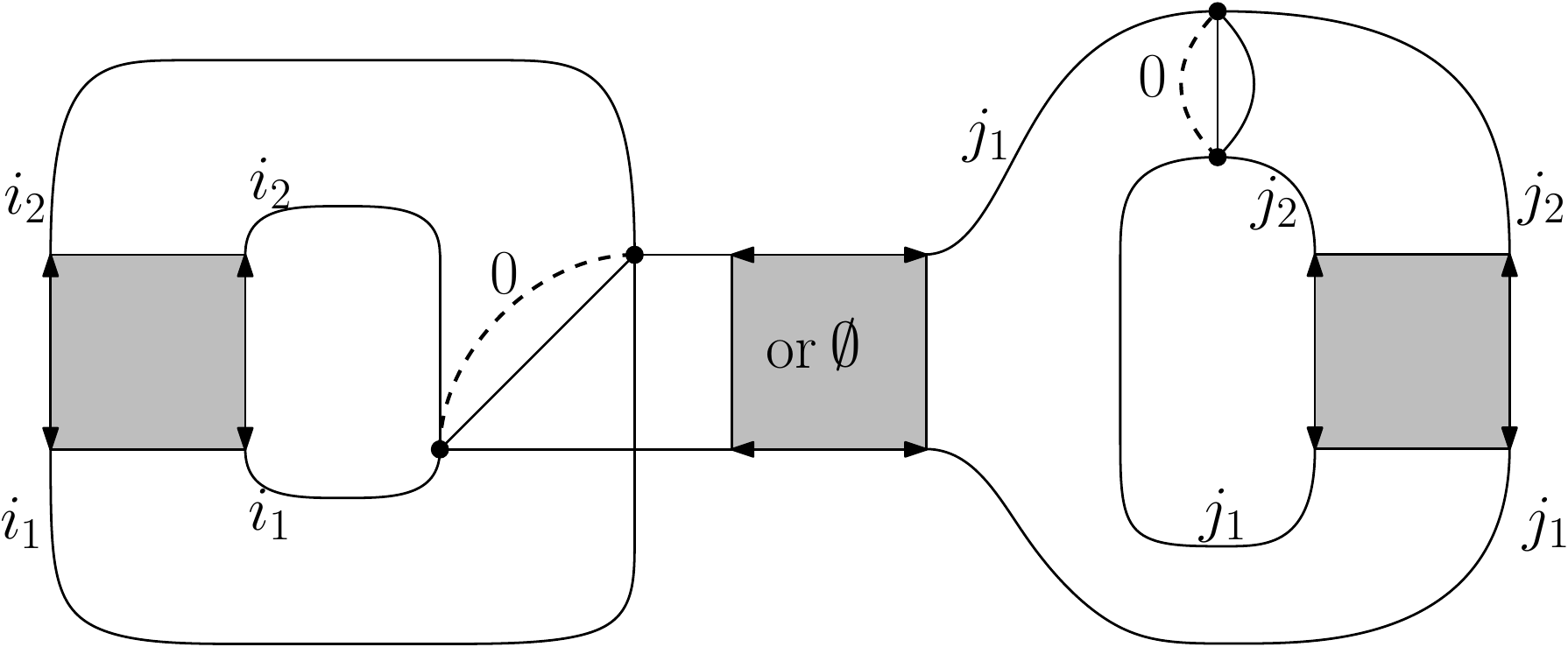} \end{array}
\end{equation}
\begin{equation}
\label{SYK_Vacuum_NNLO2}
\begin{array}{c} \includegraphics[scale=.4]{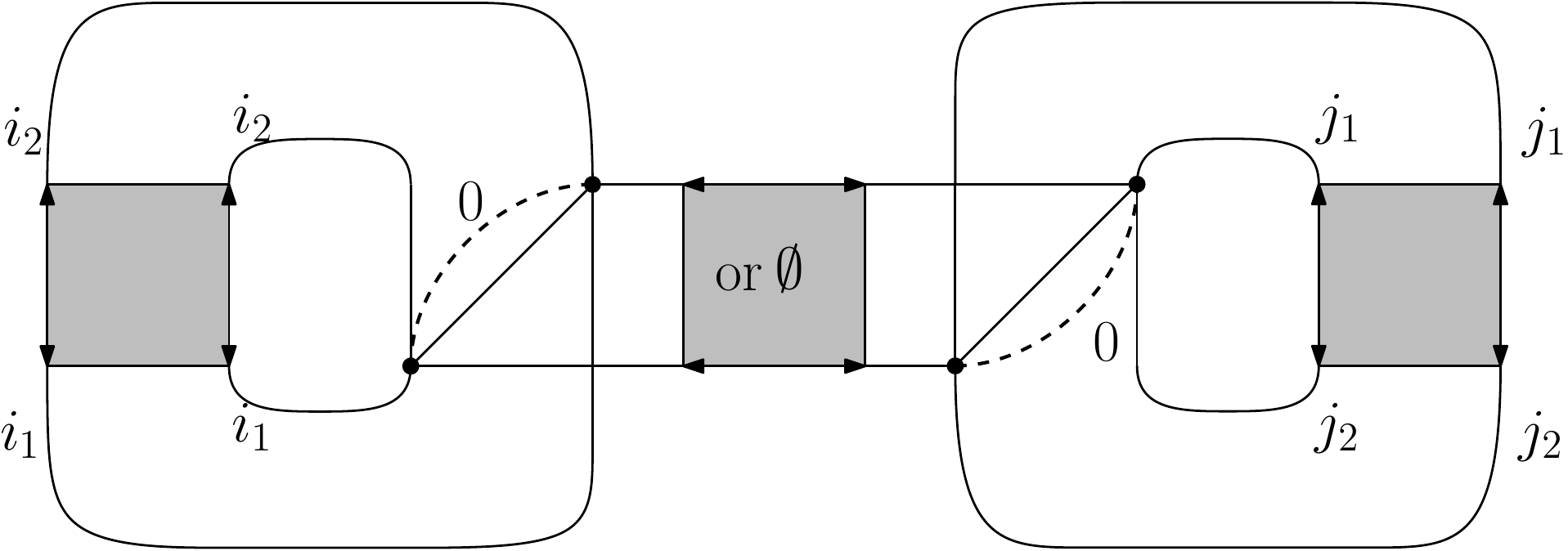} \end{array} \hspace{2cm} 
\begin{array}{c} \includegraphics[scale=.4]{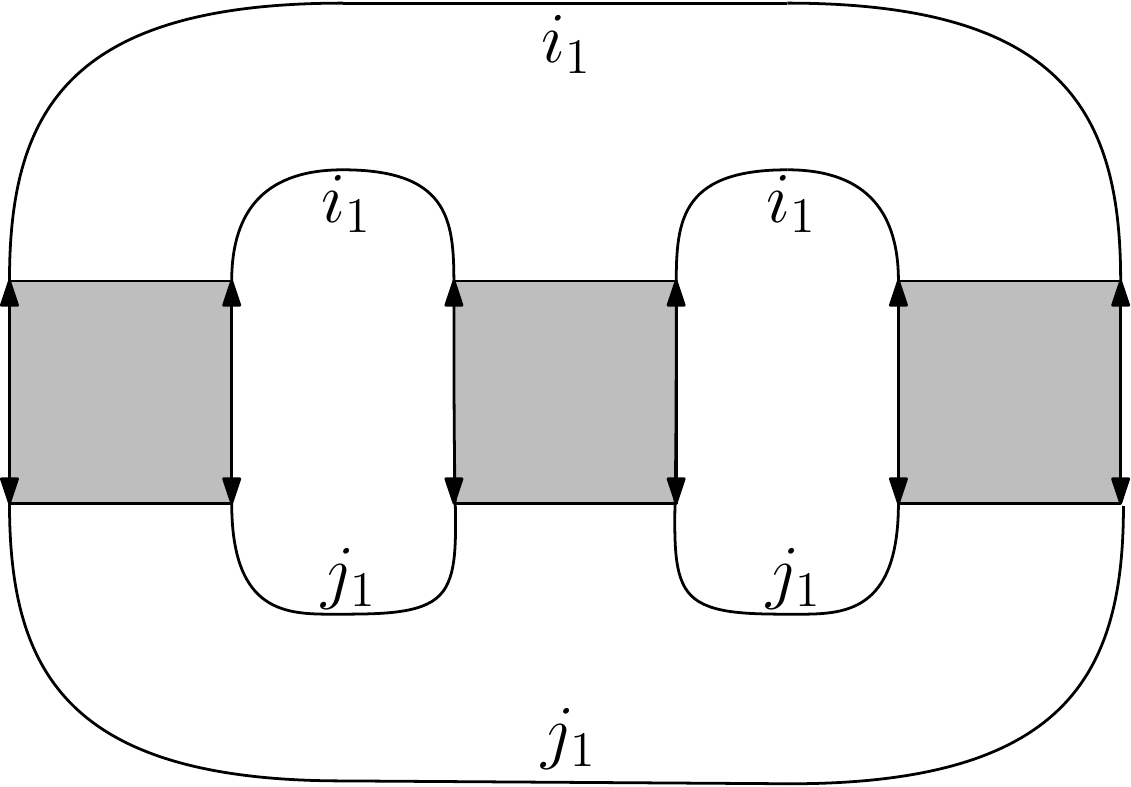} \end{array}
\end{equation}
\begin{equation}
\label{SYK_Vacuum_NNLO3}
\begin{array}{c} \includegraphics[scale=.4]{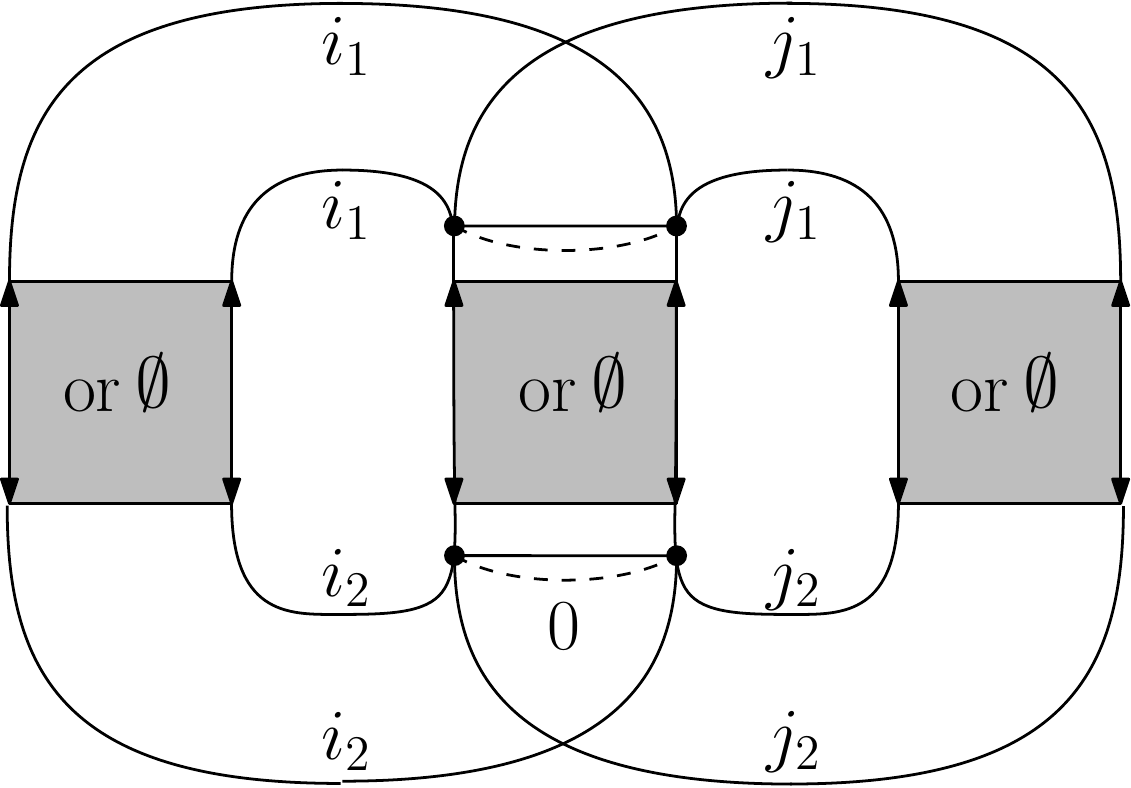} \end{array} \hspace{1cm}  
\begin{array}{c} \includegraphics[scale=.4]{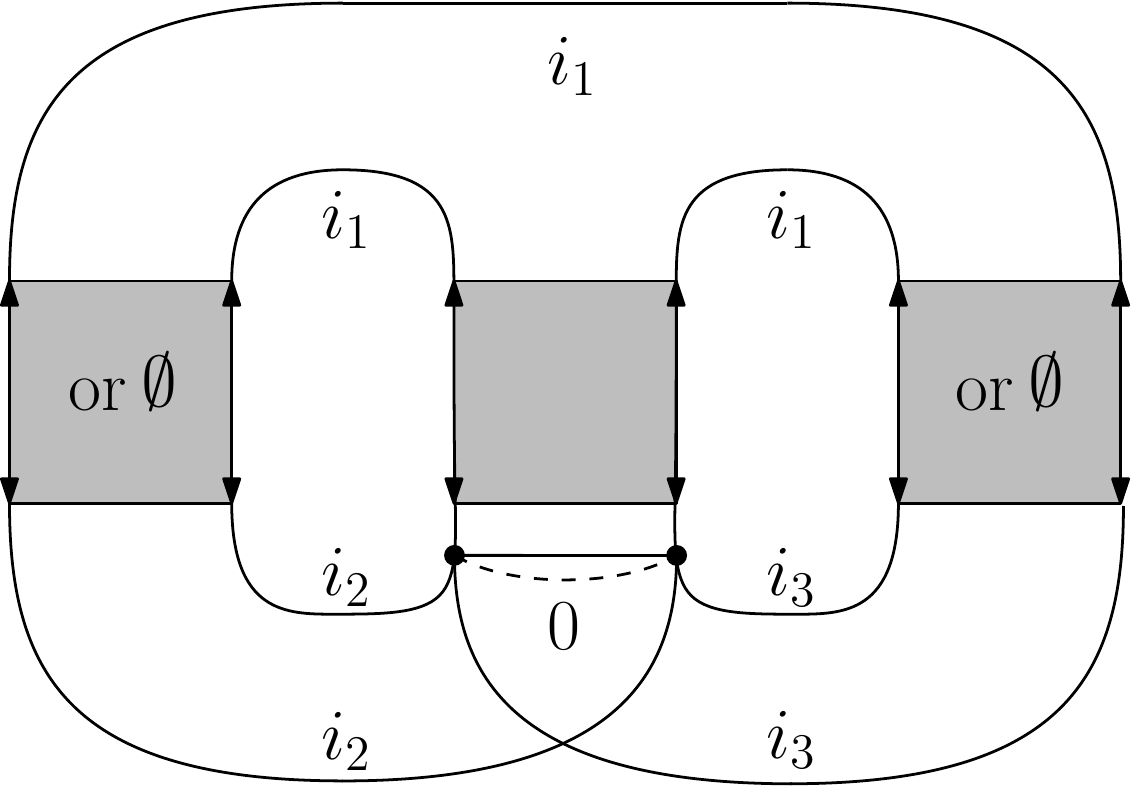} \end{array}
\hspace{1cm} 
\begin{array}{c} \includegraphics[scale=.4]{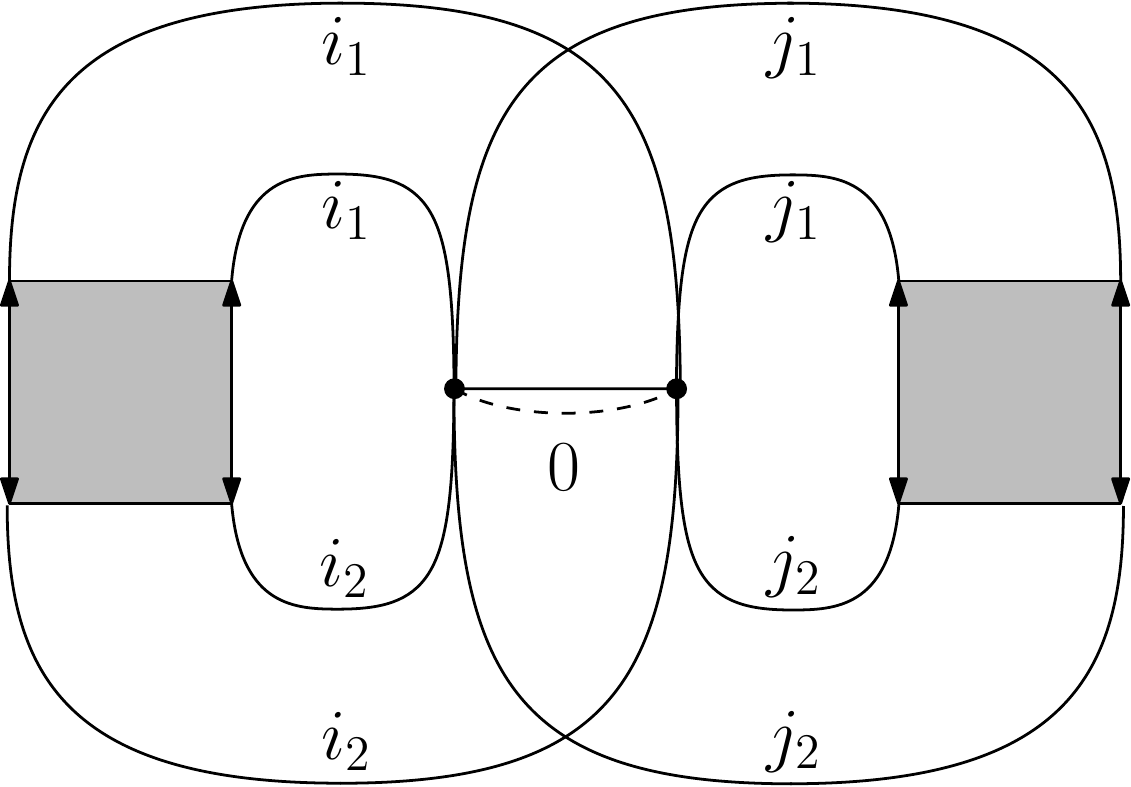} \end{array}
\end{equation}

\subsection{LO and NLO of 4-point functions}

It is easy to see that the external legs of 4-point graphs come in pairs where two legs of a pair share the same color. This gives two sets of 2-point functions, depending on whether all external legs have the same color or not,
\begin{equation}
\langle \psi_i \psi_i \psi_i \psi_i \rangle \quad \text{for $i\in\{1, \dotsc, q\}$, and} \quad \langle \psi_i \psi_i \psi_j \psi_j\rangle \quad \text{for $i\neq j$},
\end{equation}
where $\psi_i, \psi_j$ are fermions of colors $i$ and $j$. Here we have dropped the time dependence since we are only concerned with the diagrammatics. We have also left the vector indices of the fermions implicit, since they are just product of Kronecker deltas which follow easily from the graphs we will give.

There are no major diagrammatic differences between the two types of 4-point functions. We will thus treat both simultaneously.

4-point functions can be obtained by cutting two edges in a vacuum graph. They can be two edges with the same color or two different colors in $\{1, \dotsc, q\}$. If $G$ is a vacuum graph, we denote $G_{e, e'}$ the 4-point graph obtained by cutting $e$ and $e'$. Obviously, if $G_4$ is a 4-point graph, there is a (possibly non-unique) way to glue the external lines two by two, creating two edges $e, e'$,  and to thus get a vacuum graph $G$ such that $G_4 = G_{e,e'}$.

Faces of $G$ and $G_{e, e'}$ are the same except for those which go along $e$ and $e'$. When $e$ and $e'$ have distinct colors, two different faces go along them in $G$ and are thus broken in $G_{e, e'}$. When $e$ and $e'$ have the same color, there can be one or two faces along them. Therefore, the weight received by $G_{e,e'}$ reads
\begin{equation}
w_N(G_{e,e'}) = N^{\chi_0(G_{e,e'})}, \qquad \text{with} \qquad \chi_0(G_{e,e'}) = \chi_0(G) - \eta(G_{e,e'}) \leq 1 - \eta(G_{e,e'}),
\end{equation}
where $\eta(G_{e,e'}) \in\{1, 2\}$ is the number of faces broken by cutting $e$ and $e'$ in $G$.

The classification thus seems a little intricate because of the two possible values for $\eta(G_{e,e'})$. We however claim that it is sufficient to only consider the graphs $G$ with edges $e, e'$ such that
\begin{equation}
\eta(G_{e,e'}) = 2.
\end{equation}
This is obviously always the case when $e$ and $e'$ have different colors. Let us thus focus on the case where $e$ and $e'$ have the same color $i\in\{1, \dotsc, q\}$. Let $G_4$ be a 4-point graph with 4 external legs of color $i$. We claim that there is always one way to connect the external legs pairwise into two edges $e$ and $e'$ with two different faces along them. Denoting $G$ this vacuum graph, we thus interpret $G_4$ as the graph $G_{e,e'}$ with $\eta(G_{e,e'}) = 2$.

With the same notations, we have thus found that
\begin{equation}
\chi_0(G_{e,e'}) = \chi_0(G) - 2.
\end{equation}
The strategy is thus for both types of 4-point functions:
\begin{itemize}
\item use the classification of vacuum graphs which we have established: LO, NLO graphs, etc.
\item cut two edges of them such that $\eta(G_{e,e'})=2$.
\end{itemize}

In the large $N$ limit, cutting two edges in melonic graphs (such that $G_{e,e,'}$ remains connected, as well as $G_{e,e'}$ minus its edges of color 0) precisely leads to the chains introduced in \eqref{SYKChain} (one might add 2-point insertions on the external legs).

At NLO, one finds
\begin{equation} \label{NLO4Pt1A}
\begin{aligned}
&A_1 = \begin{array}{c} \includegraphics[scale=.6]{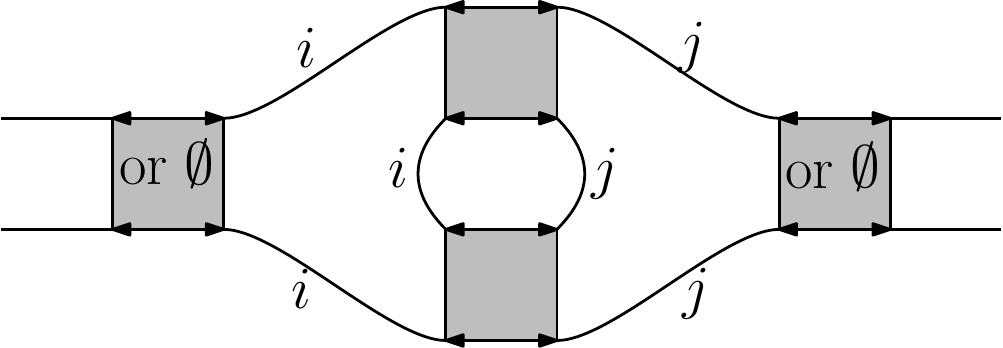} \end{array} & A_2 = \begin{array}{c} \includegraphics[scale=.6]{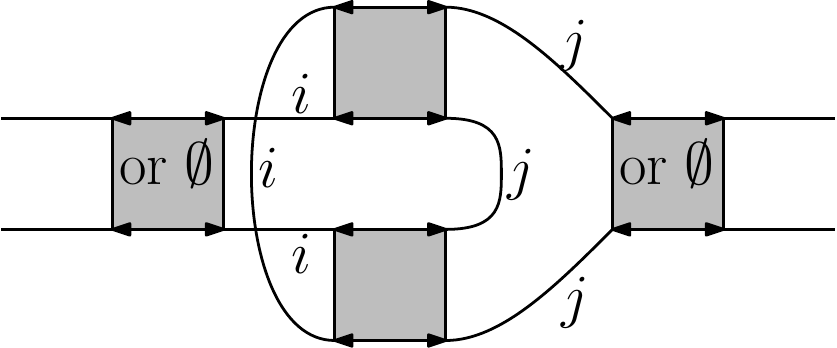} \end{array}\\
&A_3 = \begin{array}{c} \includegraphics[scale=.6]{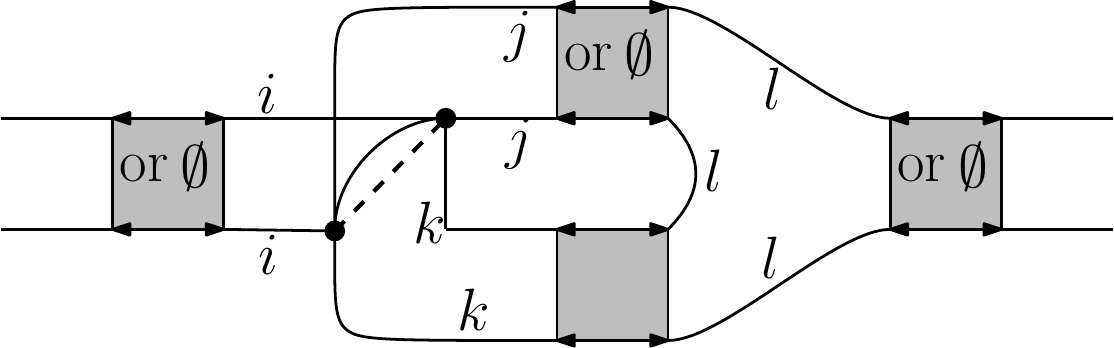} \end{array} & A_4 = \begin{array}{c} \includegraphics[scale=.6]{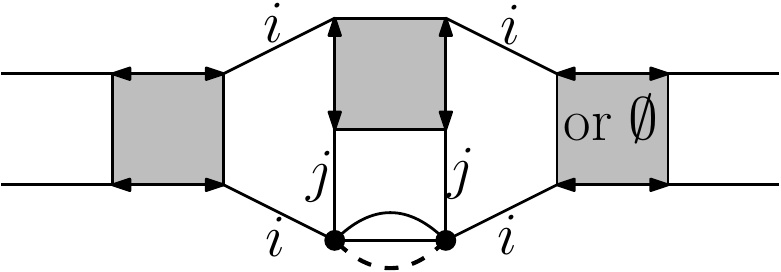} \end{array}
\end{aligned}
\end{equation}
by cutting an edge in $G_2^{\text{NLO}(1)}$ in \eqref{NLO2Pt1},
\begin{equation}
\label{NLO4Pt1B}
\begin{aligned}
&B_1 = \begin{array}{c} \includegraphics[scale=.6]{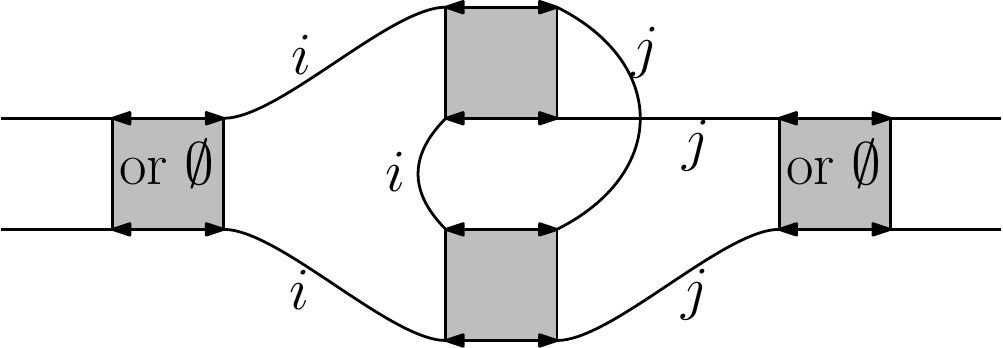} \end{array} & B_2 = \begin{array}{c} \includegraphics[scale=.6]{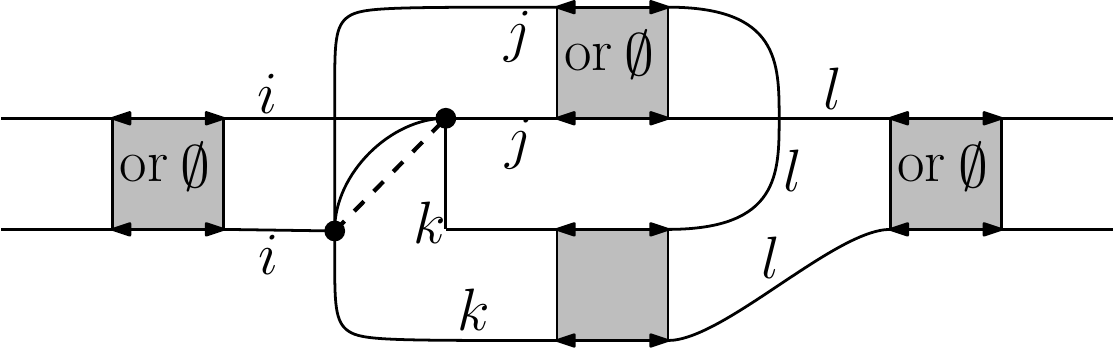} \end{array}\\
&B_3 = \begin{array}{c} \includegraphics[scale=.6]{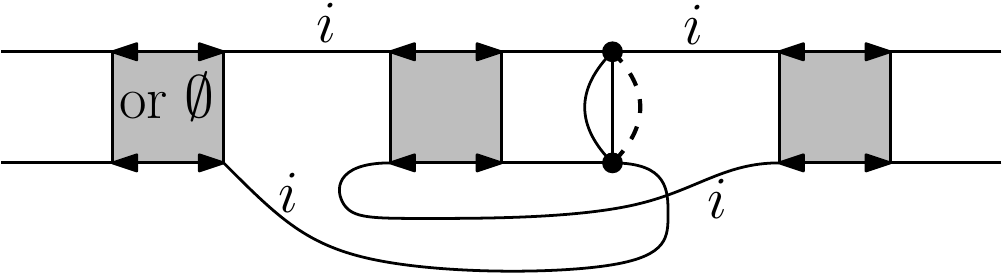} \end{array} & 
\end{aligned}
\end{equation}
by cutting an edge in $\tilde{G}_2^{\text{NLO}(1)}$ in \eqref{NLO2Pt1},
\begin{equation}
\label{NLO4Pt1C}
C_1 = \begin{array}{c} \includegraphics[scale=.6]{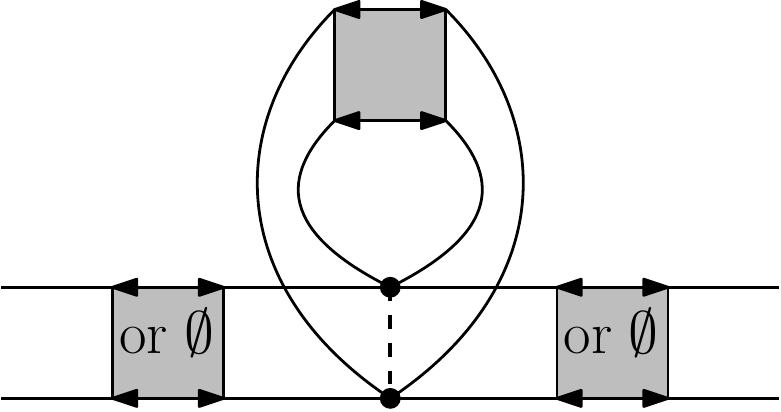} \end{array} \qquad C_2 = \begin{array}{c} \includegraphics[scale=.6]{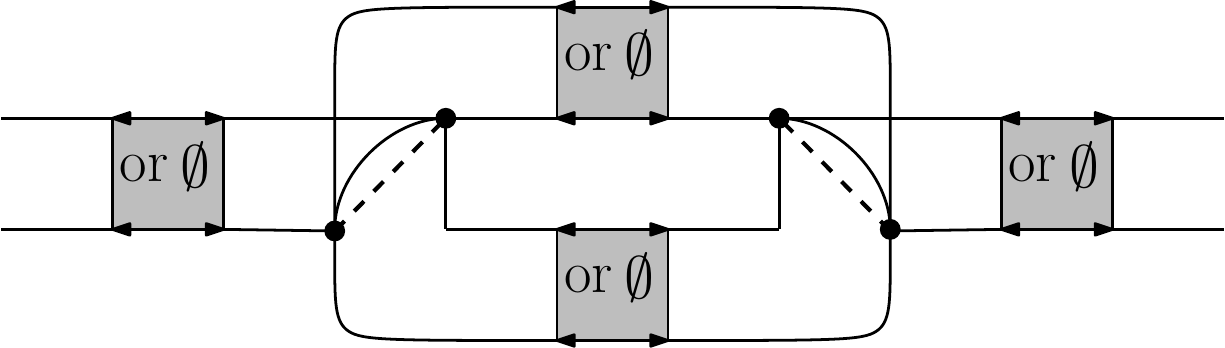} \end{array}
\end{equation}
by cutting an edge in $G_2^{\text{NLO}(2)}$ in \eqref{NLO2Pt2},
\begin{equation}
\label{NLO4Pt1D}
\begin{aligned}
&D_1 = \begin{array}{c} \includegraphics[scale=.6]{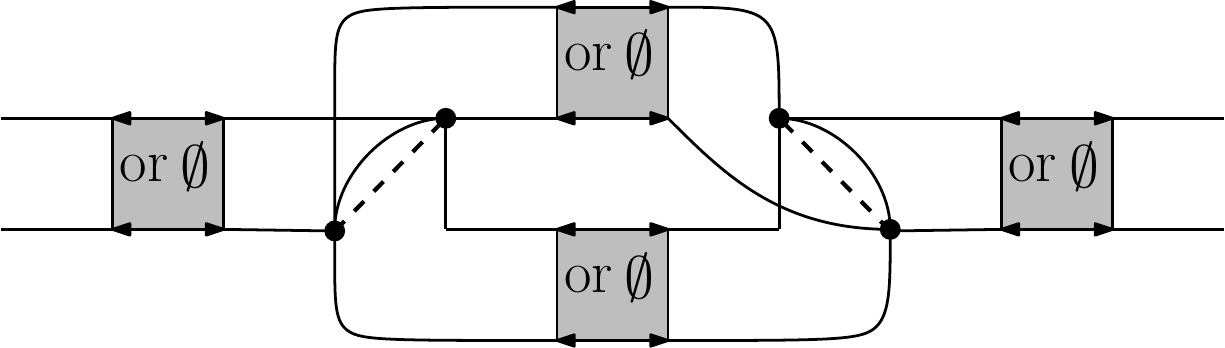} \end{array} & D_2 = \begin{array}{c} \includegraphics[scale=.6]{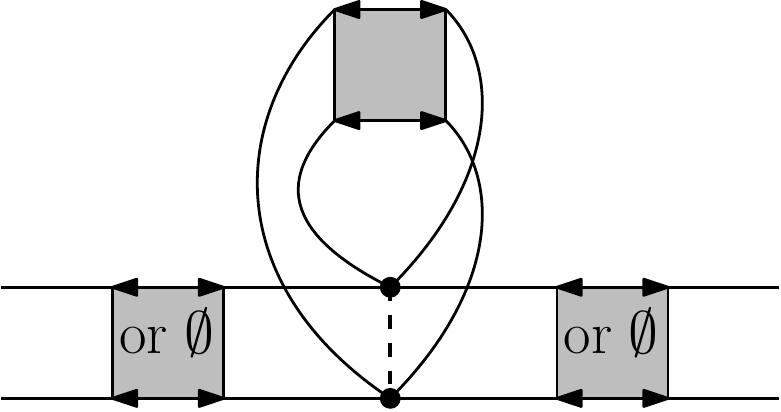} \end{array}
\end{aligned}
\end{equation}
by cutting an edge in $\tilde{G}_2^{\text{NLO}(2)}$ in \eqref{NLO2Pt2}, and finally the two following families
\begin{equation}
\label{NLO4Pt1E}
\begin{array}{c} \includegraphics[scale=.6]{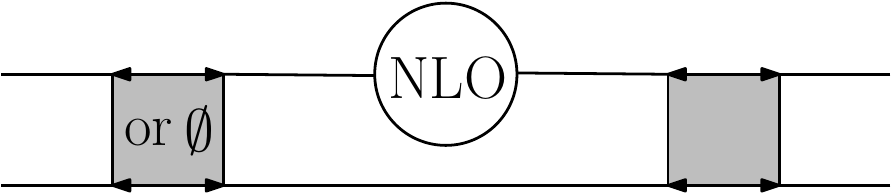} \end{array} \qquad
\begin{array}{c} \includegraphics[scale=.6]{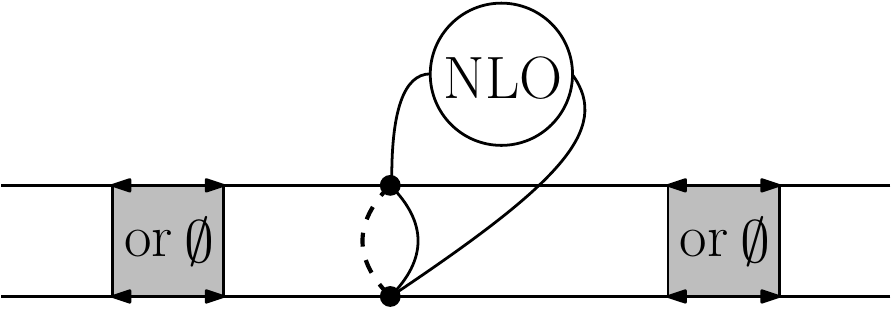} \end{array} 
\end{equation}
obtained by performing a 2-point insertion of the NLO 2-point function into the LO 4-point chains.

The above description avoids redundancies. Notice that the colors are important. For instance, by specializing the middle chains in $A_2$ and $A_4$ to have a single pair of vertices, the same graph is obtained but with different colorings.

\section{Diagrammatics of the Gurau-Witten model} \label{sec:GW}

We focus in this section on the Gurau-Witten model whose action is \cite{Gurau, Witten}
\beqa
\label{actGW}
S=\int dt \left( 
\frac{\imath}{2}\sum_{f=1}^4 \psi^f\frac{d}{dt}  \psi^f
+\frac \lambda {N^{\frac {D(D-1)} 4}} \psi^1\psi^2\psi^3\cdots \psi^q
\right)
\eeqa
Note that the $q=D+1$ fields above $\psi^{f}$, $f=1,\ldots, q$ are now rank $q-1$ tensor fields. The notation $\psi^1\psi^2 \dotsm \psi^q$ in the interaction has a specfic pattern of index contraction. The tensor $\psi^i$ has $q-1$ indices $(n_{i i-1}, \dotsc, n_{i1}, n_{iq}, \dotsc, n_{i i+1})$ where each $n_{ij}$, $i\neq j$, has range $n_{ij} = 1, \dotsc, N$. The contraction in the interaction is defined by setting $n_{ij} = n_{ji}$, which identifies one index of $\psi^i$ with one index of $\psi^j$.

\begin{figure}
\begin{center}
\includegraphics[scale=0.9]{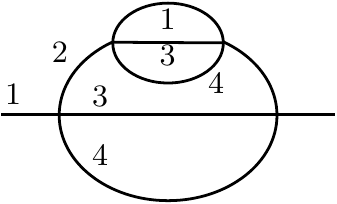}
\caption{\label{melonGW}A melonic graph of the Gurau-Witten model.}
\end{center}
\end{figure}

The Feynman graphs obtained through perturbative expansion are stranded graphs where each strand represents the propagation of an index $n_{ij}$, alternating stranded edges of colors $i$ and $j$. However, since no twists among the strands are allowed, one can easily represent the Feynman tensor graphs as standard Feynman graphs with additional colors on the edges. Those colors are just the labels $f$ of the tensor fields $\psi^f$. A vertex has degree $q$ and is incident to exactly one edge of each color in $\{1, \dotsc, q\}$. An example of such a Feynman diagram is given in Figure \ref{melonGW} for $q=4$.
\emph{The graphs are thus exactly those of the colored SYK model of sections \ref{sec:GeneralizedSYK}, \ref{sec:SYK} at fixed couplings $j$.} However, the scaling with $N$ is quite different.

The partition function pertubatively expands as a $1/N$ expansion, in which each graph $G$ receives a weight 
\begin{equation}
\label{GWdeg}
w_N(G) = N^{\chi(G)} \qquad \text{with $\chi(G)= F(G) - \frac {(q-1)(q-2)} 4V(G)$},
\end{equation}
in which we denoted $V(G)$ the number of vertices of $G$, which is also twice its number of edges of any color $i$. Let us explain briefly the origin of $F(G)$. Each graph indeed receives a free sum from $1$ to $N$ for each cycle alternating the colors $i<j$.  They are called faces of colors $ij$ and we denote $F_{ij}(G)$ their numbers. The total number of faces is 
\begin{equation}
F(G) = \sum_{\substack{i,j=1\\ i<j}}^{q} F_{ij}(G).
\end{equation}
Instead of $\chi(G)$ one can introduce the reduced degree 
\begin{equation}
\delta(G)=q-1-\chi(G)
\end{equation}
as it is commonly used in the random tensor literature
\begin{equation} \label{ReducedDegree}
  \delta(G)= q-1+ \frac{(q-1)(q-2)} 4 V(G) -F(G),
\end{equation}
for a connected graph. It is proportional to Gurau's degree $\omega(G)$ 
\begin{equation} \label{OmegaDelta}
\frac{2}{(q-3)!} \omega(G)= \delta(G),
\end{equation}
which is itself also defined as $\omega(G) =\sum_\cJ g(G_\cJ)$, where $g(G_\cJ)$ are the genera of the so-called jackets (see \cite{Gurau} and some details below).

Gurau's theorem on the $1/N$ expansion of tensor models \cite{Complete1/N} ensures that the reduced degree is positive  and vanishes for melonic graphs only, so that $\chi$ is bounded,
\begin{equation}
\chi(G)= F(G) - \frac {(q-1)(q-2)} 4V(G) \leq \begin{cases} q-1 & \text{if $G$ is a vacuum graph,}\\
0 & \text{if $G$ is a 2-point graph.} \end{cases}
\end{equation}
The case of 4-point graphs will be discussed below.

There is a complex version of this model presented in \cite{Gurau}. Then, the Feynman graphs of the perturbative expansion are bipartite, but this is not always true anymore in the real model \eqref{actGW}. When the graphs are not bipartite, the jackets can be non-orientable surfaces and thus have half-integer genera, like in the $O(N)$ and multi-orientable random tensor models \cite{CT, mosigma}.

\subsection{2-point functions: LO, NLO and so on} \label{sec:2PtGW}

As already mentioned, a 2-point graph is obtained by cutting a line from a vacuum graph. If $G$ is the vacuum graph and $e$ the edge being cut, we denote $G_e$ the 2-point graph. The other way around, there is a unique way to glue the external lines of a 2-point graph together to form a vacuum graph, so that all 2-point graphs can be thought of as a graph $G_e$. We say that $G$ is the closure of $G_e$.

The powers of $N$ in $G$ and $G_e$ differs by $q-1$,
\begin{equation}
\chi(G_e) = \chi(G) - (q-1),
\end{equation}
since $q-1$ faces are broken by cutting $e$. 
For a 2-point graph $G_e$, we will always refer to its reduced degree as the reduced degree $\delta(G)$ of its closure $G$.

Dominant graphs are melonic graphs (see Figure \ref{melonGW}) which are the graphs of vanishing Gurau's degree $\omega(G) = 0$. This is the same for the SYK model, for the colored tensor model, for the MO tensor model and even generic tensor models \cite{uncolored}.

\subsubsection{Results}

The classification of 2-point colored graphs with respect to Gurau's degree has been performed in \cite{GS} for bipartite graphs (e.g. coming from a complex tensor model) and we thus only need to incorporate non-bipartite graphs.

The main result of this section is that NLO, 2-point graphs are bipartite, and thus given by the Gurau-Schaeffer classification \cite{GS, Gurau}. They have reduced degree $\delta(G) = q-3$ and exponent $\chi(G) = 2$ and the corresponding $G_e$ are \cite{GS},
\begin{equation} \label{NLO2PtGW}
\begin{array}{c} \includegraphics[scale=.45]{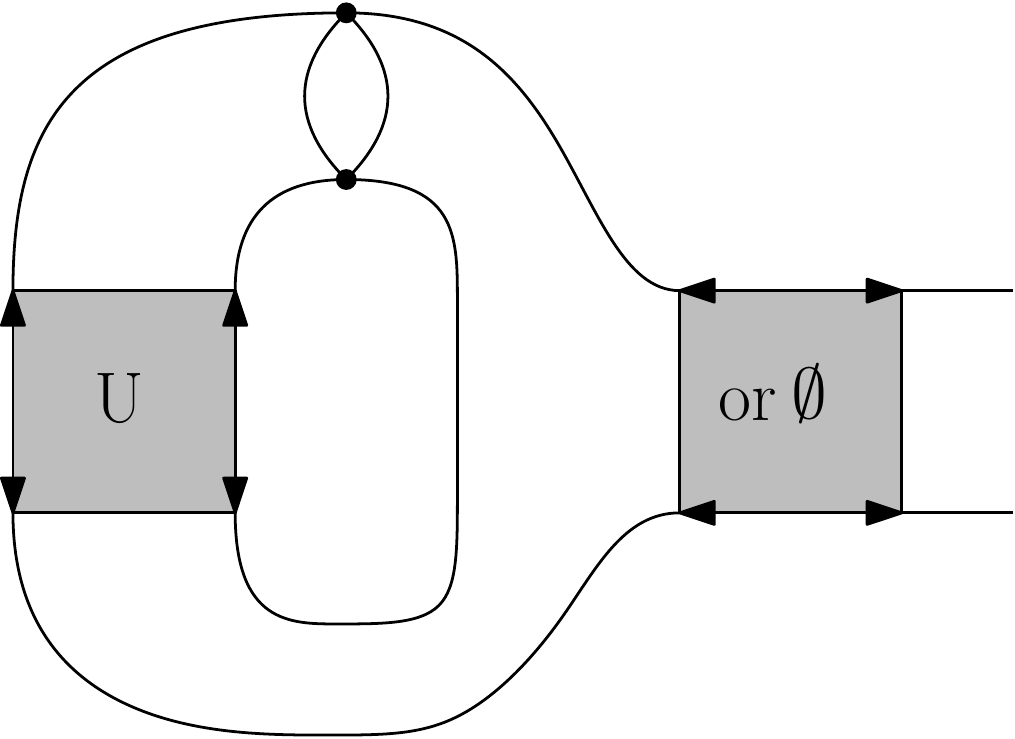} \end{array} \hspace{2cm} \begin{array}{c} \includegraphics[scale=.45]{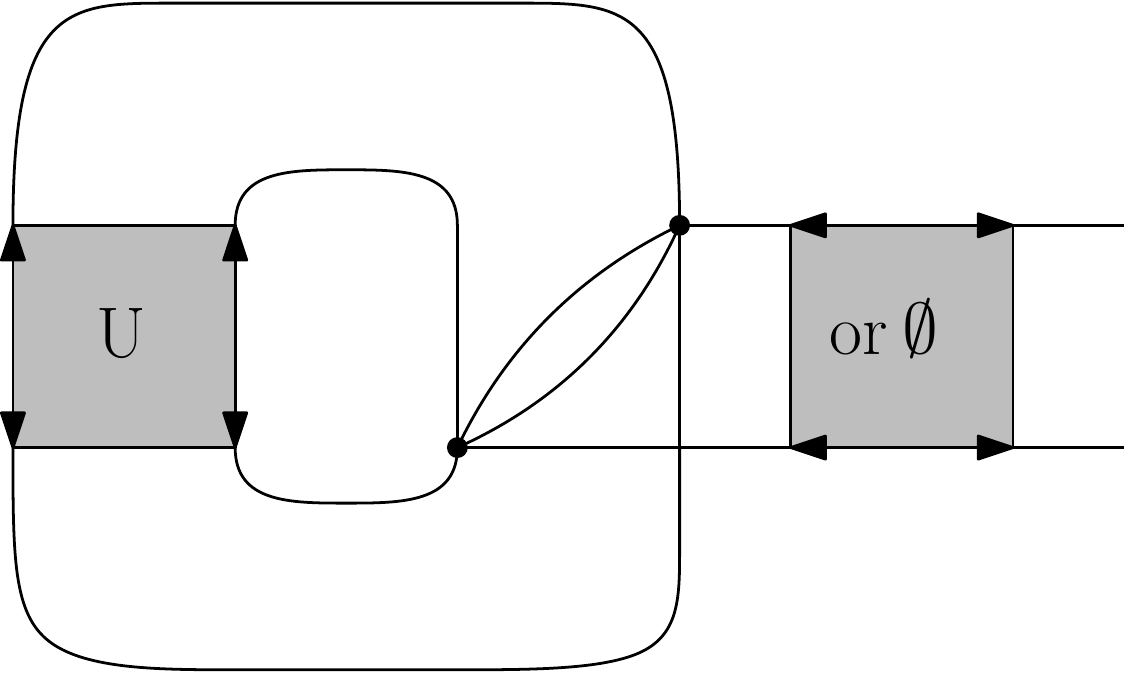} \end{array}
\end{equation}
Here the box with a U inside is an \emph{unbroken} chain, corresponding to any 4-point graph of the form
\begin{equation} \label{UnbrokenDef}
\begin{array}{c} \includegraphics[scale=.6]{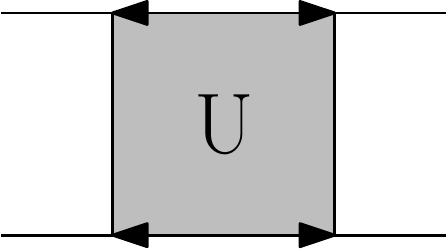} \end{array} = \left\{ \begin{array}{c} \includegraphics[scale=.6]{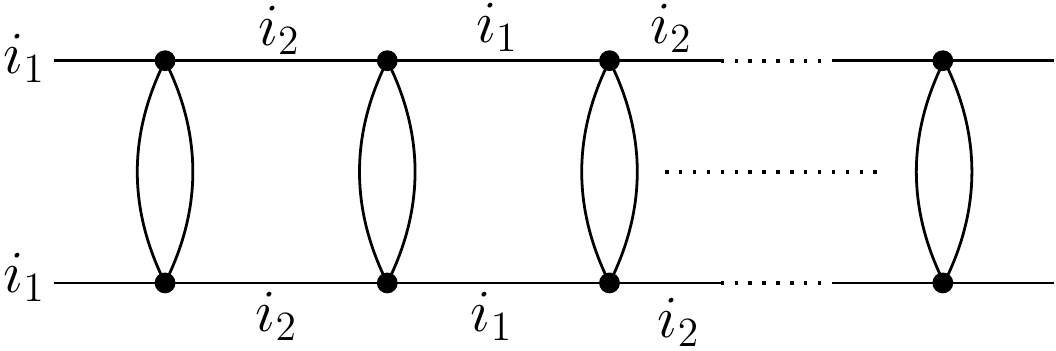} \end{array} ; i_1, i_2 \in\{1, \dotsc, q\} \right\}
\end{equation}
where the same two colors $i_1, i_2$ alternate along the chain (its right hand side edges are of color $i_1$ or $i_2$ depending on the parity of the number of pairs of vertices) and it is allowed to have a single pair of vertices. A box without a U inside is any chain, i.e. any sequence of colors can be found along the chain.

Non-bipartite graphs first appear at NNLO, with $\chi(G)=1$ ($\delta(G)=q-2$).  
The only known examples are the following,
\begin{equation} \label{NNLO2PtGW}
\begin{array}{c} \includegraphics[scale=.45]{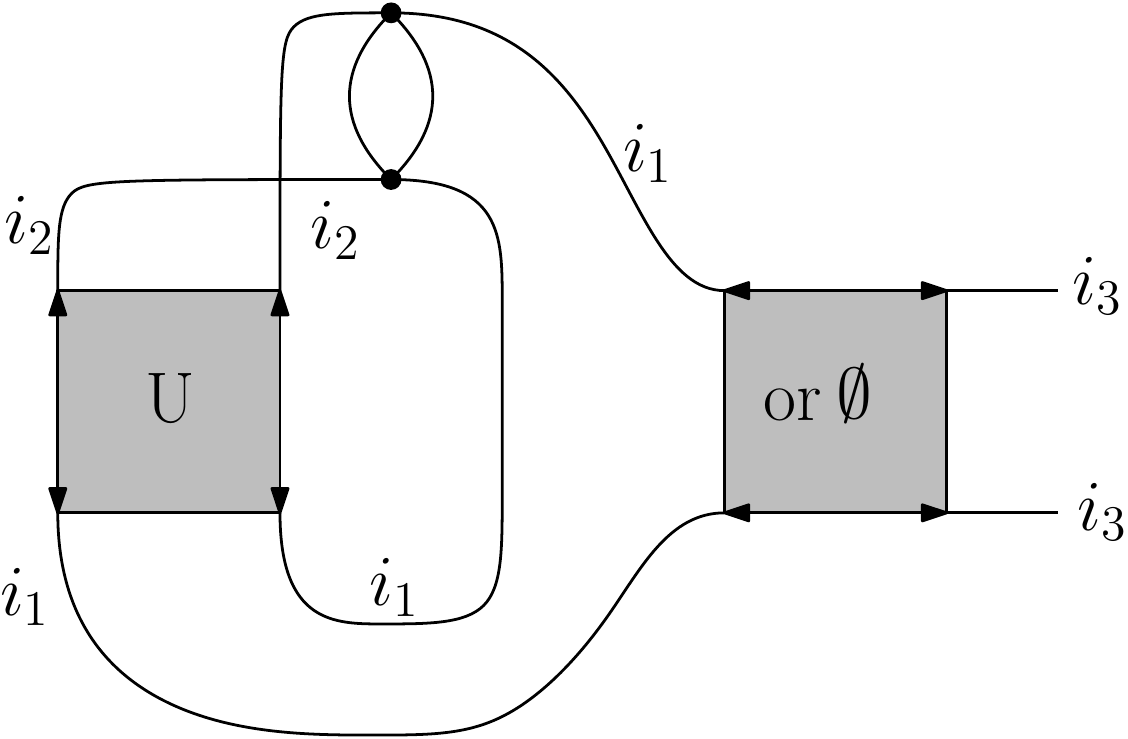} \end{array} \hspace{2cm} \begin{array}{c} \includegraphics[scale=.45]{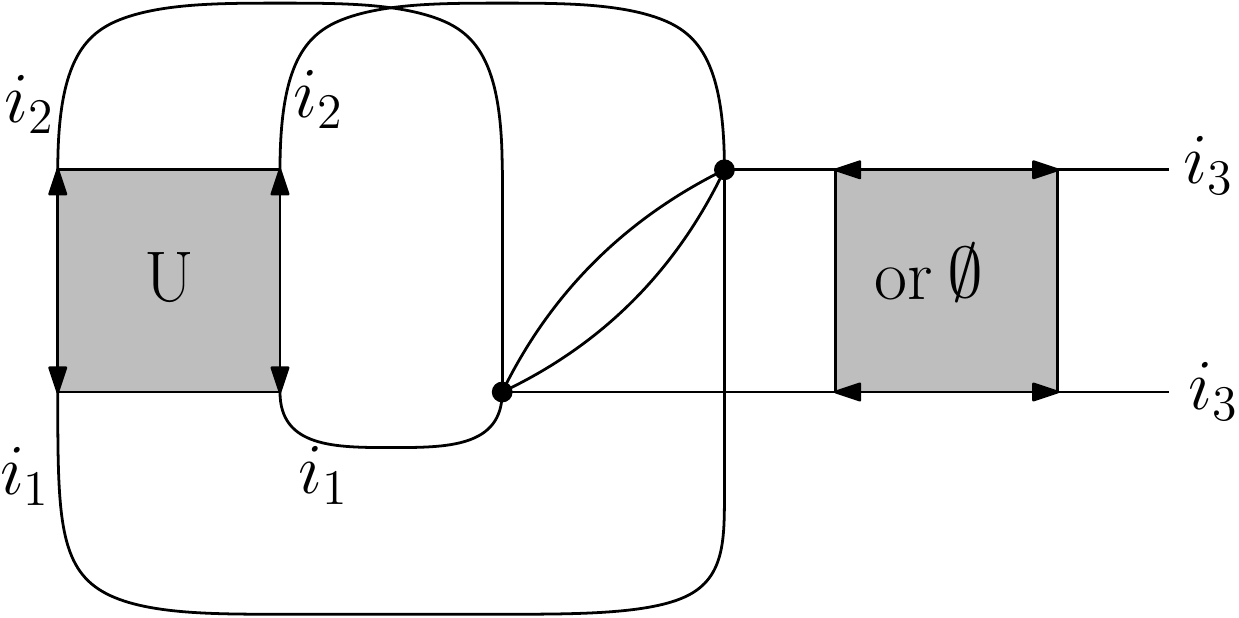} \end{array}
\end{equation}
The graphs in \eqref{NLO2PtGW} and \eqref{NNLO2PtGW} look very much like the NLO 2-point graphs \eqref{NLO2Pt1} and \eqref{NLO2Pt2} of the colored SYK model (with unbroken chains being just SYK chains). However in the SYK case the bipartite and non-bipartite graphs have the same exponent of $N$, which is not the case here.

\subsubsection{Proof}

We will prove that non-bipartite graphs are at best NNLO, i.e. they have reduced degree $\delta(G) \geq q-2$. The proof proceeds by induction on the number of colors $q$. Since our proof uses classical tools of random tensor/colored graph theory, we will use the more common variable
\begin{equation}
D = q-1
\end{equation}
so that the graphs are edges carrying the colors $1, \dotsc, D+1$, and we want to prove that $\delta(G) \geq D-1$ for non-bipartite graphs.

We start with $D=3$, i.e. $q=4$. We recall \cite{Gurau, Witten} that the degree of a $4$-colored graph $G$ is defined as
\begin{equation}
\label{SumJaq}
\omega(G)=\sum_\cJ g(G_\cJ),
\end{equation}
where there are $3$ ribbon graphs $G_\cJ$ called \emph{jackets} and $g(G_\cJ)$ is the genus of $G_\cJ$. Each $G_\cJ$ has all the vertices and edges of $G$, but only a subset of faces determined by a permutation up to cyclic ordering and orientation reversal\footnote{The three permutations are here $\cJ = (1234), (1243), (1423)$. The faces of $G_{\cJ}$ are then the bicolored cycles with colors $(\cJ^i(1) \cJ^{i+1}(1))$.}. From \eqref{ReducedDegree}, \eqref{OmegaDelta}, we know that $\omega(G)$ is an integer. This implies that jackets with half-integral genera come in pairs.

Let us now show that if $G$ is non-bipartite, $\omega(G)\ge2$. Assume that $G$ has a planar jacket, i.e. $g(G_\cJ)=0$ for some $\cJ$. Moreover, all faces of $G_\cJ$ have even length since they are bicolored. As a result, $G_\cJ$ is a planar ribbon graph with faces of even length. A classical result in graph theory then ensures that $G_\cJ$, hence $G$ is bipartite. Therefore, a non-bipartite graph has no planar jackets. To minimize the degree, it must have two jackets with genera 1/2 and one of genus 1, i.e. $\omega(G)\geq2$. Since $\delta(G) = 2 \omega(G)/(D-2)!$, we conclude that $\delta(G) \geq D-1$ at $D=3$.
  
As the previous argument does not work for larger values of $D$, we now perform an induction. To do so, we will distinguish a color $i\in\{1, \dotsc, D+1\}$ and control the number of faces of colors $(ij)_{j\neq i}$ using the results we derived for the SYK model, where the role of the color $0$ is played by the color $i$. The obstacle is however that in the SYK model, removing the color $0$ does not disconnect the graph, while removing the color $i$ in the present model typically does disconnect the graph. We will show how to overcome that obstacle so that indeed adding the color $i$ to a graph with $D$ colors has the same diagrammatic effect as averaging over disorder in the SYK.

Therefore, we introduce a new operation on the graph, known as the 1-dipole contraction \cite{Complete1/N}. If $G$ is connected, deleting every edge of color $i\in\{1,\cdots, D+1\}$, one obtains a graph with $D$ colors only, which we denote $G_{\hat i}$ and whose connected components are called \emph{bubbles}. If an edge of color $i$ separates two distinct bubbles, a 1-dipole contraction can be performed as follows. One first contracts the edge as described in (\ref{Contraction}). Deleting the resulting vertex one is left with a set of un-contracted half-edges, two for each color, which are then contracted while respecting the colors, as follows
\begin{equation}
\label{Contraction2}
\begin{array}{c} \includegraphics[scale=.5]{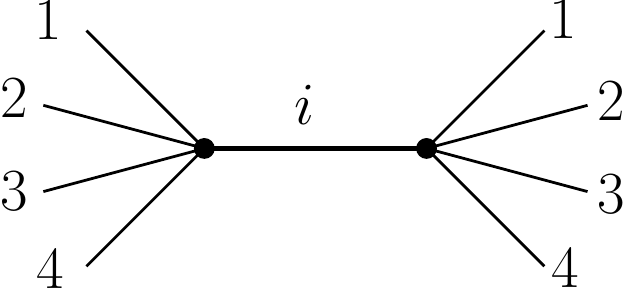} \end{array} \qquad \underset{/i}{\to} \qquad \begin{array}{c} \includegraphics[scale=.5]{Edge0Contracted.pdf} \end{array} \qquad {\to} \qquad\begin{array}{c} \includegraphics[scale=.5]{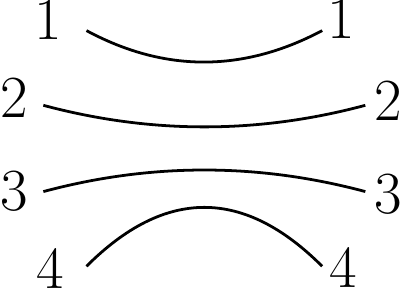} \end{array}
\end{equation}
Crucially, a 1-dipole contraction does not change the degree (nor the topology of the underlying $D$-dimensional cell complex), as shown in \cite{Complete1/N} (Lemma 2), and also preserves (non-)bipartiteness\footnote{
Clearly, a 1-dipole contraction of a bipartite graph is still bipartite. The other way around, consider a (connected) bipartite bubble and a $D$-cut of that bubble, i.e. a set of $D$ edges $e_1, \dotsc, e_D$ with the colors $1, \dotsc, D$ which can be cut to create several connected components. Denote $v_k$ the white vertex of $e_k$ and $v'_k$ its black vertex for $k=1, \dotsc, D$. All vertices $(v_k)$ are in the same connected component after the cut is performed. Indeed, since each color is incident exactly once on each vertex, except the color $k$ on $v_k$ and $v'_k$ for all $k$, there is a path alternating the colors 2 and 1 starting from $v_1$ and which must end on $v_2$ or $v'_2$. Since the path is alternating, it actually ends on $v_2$, and similarly with the others $v_k$. Therefore all $(v_k)$ are in one connected component and all $(v'_k)$ are in another after the cut. It is then clear that one can perform a 1-dipole insertion, by inserting a new edge with color $D+1$, while preserving bipartiteness.}.

A series of 1-dipole contractions on edges of a fixed color $i\in\{1, \dotsc, D+1\}$ can turn a connected graph $G$ into a new graph $T$ with a single bubble $T_{\hat{i}}$, i.e. such that the graph $T_{\hat{i}}$, having all edges of color $i$ removed, is connected. Indeed, we perform a 1-dipole contraction for some edge of color $i$ between two connected components of $G_{\hat i}$. The result has one connected component less and we can perform a 1-dipole contraction on another edge between two of the remaining connected components and so on, until only one connected component $T_{\bar i}$ remains.

The reduced degree of $T$ is that of $G$, $\delta(G)=\delta(T)$ and moreover
\begin{equation} \label{SplitDegree}
\delta(G)=\delta(T)=\delta_i(T) + \delta(T_{\hat i}),
\end{equation}
where $\delta(T_{\hat i})$ is the degree of the bubble as a colored graph with $D$ colors,
\begin{equation}
\delta(T_{\hat i}) = D-1 + \frac{(D-1)(D-2)}{4} V(T) - \sum_{\substack{j, k\neq i\\ j<k}} F_{jk}(T),
\end{equation}
and $\delta_i(T)$ is 
\begin{equation}
\delta_i(T) = D - F_i(T) + (D-1)( V(T)/2 - 1).
\end{equation}
Note that $\delta_i(T)$ can be interpreted as a reduced degree in the sense of the SYK model, that is it only takes into account the number $F_i(T)$ of faces of colors $(ij)_{j\neq i}$ for fixed $i$ (in the SYK model, $i=0$). Indeed, since $T$ has a unique bubble after removing the color $i$, it can be interpreted as a graph from the SYK model with $i$ replacing $0$. Moreover, we recall that in the SYK model, the reduced degree can be defined as $D - \chi_0(T) = D + (D-1)V(T)/2 - F_0$ (see \eqref{SYKdeg} with $q=D$ in this case). Here $\delta_i(T)$ only differs from the latter by a constant term $-(D-1)$.

We can therefore use the induction hypothesis on $\delta(T_{\hat{i}})$ and the results of Section \ref{sec:2Pt} for $\delta_i(T)$. Suppose $G$ is a non-bipartite graph with $D+1>4$ colors, and suppose we have proven that non-bipartite graphs with $D$ colors have a reduced degree $\delta\ge D-2$. We know from section \eqref{sec:2Pt} that if $\delta_i(T)=0$, then $T_{\hat i}$ is melonic and the edges of color $i$ must connect its canonical pairs of vertices, therefore $T$ is also melonic and so is $G$. In particular it is bipartite, which is excluded, so that $\delta_i(T)\ge 1$.

If $\delta(T_{\hat{i}})\geq D-2$, then there is nothing to prove. From the induction hypothesis (telling us that graphs with $\delta(T_{\hat{i}})< D-2$ are bipartite and thus classified in \cite{GS}), we know that we have to look at the cases where $\delta(T_{\hat{i}}) =0, D-3$.

When $\delta(T_{\hat{i}})=D-3$, only the case $\delta_i(T) = 1$ is non-trivial. For $\delta_i(T)=1$, we showed in the previous section that $T$ is an NLO SYK graph as in (\ref{NLO2Pt1}) and (\ref{NLO2Pt2}), with color $i$ connecting the canonical pairs, as was the color 0 then. As furthermore $\delta(T_{\bar i})=D-3$, we know from the induction hypothesis that $T_{\hat i}$ is bipartite, which leaves only the cases on the left of (\ref{NLO2Pt1}) and (\ref{NLO2Pt2}). But then $T$ itself is bipartite, which is excluded as the inverse of (\ref{Contraction2}) preserves bipartiteness.

Only the case $\delta(T_{\hat{i}})=0$, i.e. $T_{\hat{i}}$ is melonic, remains to be investigated. We will do so by proving the more general property (P): for a graph $G$ with $D+1$ colors having a melonic bubble, either there exists a color $k$ and a bubble $B_{\hat{k}}$ such that $\delta(B_{\hat k})>0$, or $\delta(G)=0$, or $\delta(G)\ge D-1$. This way, if $T_{\hat{i}}$ is melonic and $G$ is not melonic, then either $\delta(G) \geq D-1$ as desired, or we can choose another color $k$ so that $\delta(T_{\hat k})>0$. From the induction hypothesis, $\delta(T_{\hat k}) \geq D-3$, which is the case we already dealt with.


We prove the property (P) inductively on the number of vertices. It is obvious if $G$ only has two vertices as it is necessarily melonic. Consider a larger graph $G$ and the melonic bubble $B_{\hat i}$ with $D$ colors. Recall that $D$-colored melonic graphs are defined as the recursive insertion of pairs of vertices connected by $D-1$ edges as on the left of (\ref{ProofMelo}), and in particular always contain such a pair of vertices. In $G$, the color $i$ can be as any one of the cases below in (\ref{ProofMelo}): $i)$ between the vertices of the pair, $ii)$ as two edges reaching another connected component so that when cutting those two edges, $G$ separates into two connected components, $iii)$ or as two edges which do not separate $G$. 
\begin{equation}
\label{ProofMelo}
\begin{array}{c} \includegraphics[scale=.5]{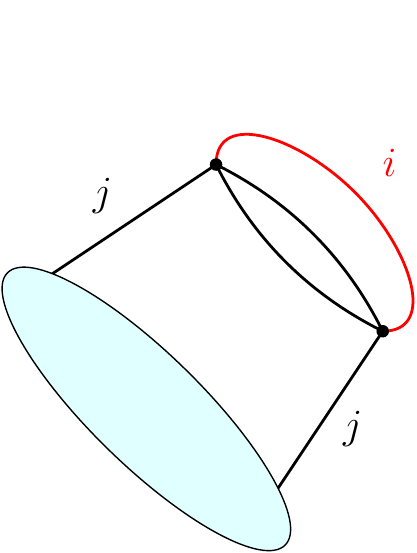} \end{array} \hspace{2cm} \begin{array}{c} \includegraphics[scale=.5]{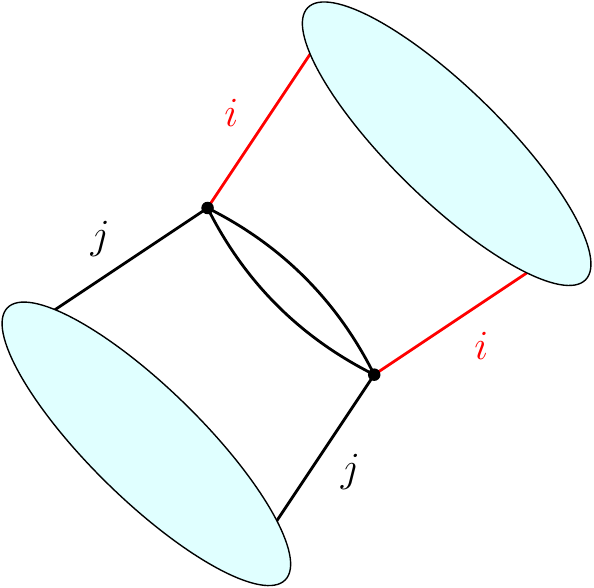} \end{array} \hspace{1.8cm}\begin{array}{c} \includegraphics[scale=.5]{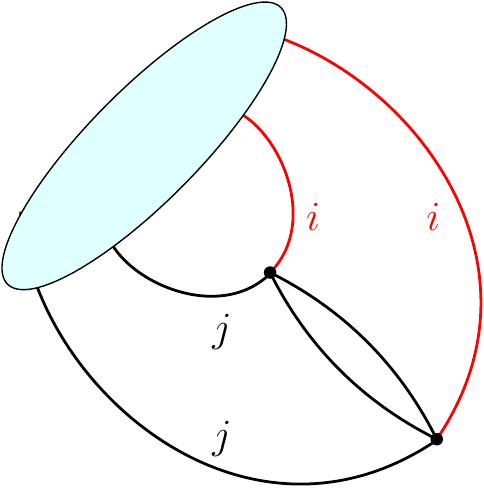} \end{array}
\end{equation}
In the case $i)$, by deleting the two vertices of the pair and reconnecting the two pending edges of color $j$, one recovers a smaller graph to which the property (P) applies. As inserting a pair of vertices connected by $D$ edges does not change neither $\delta(G)$ nor $\delta(B_{\hat k})$, (P) applies to $G$ too. In the case $ii)$, one may cut the two edges of color $i$ and reconnect the pending half-edges so that an edge of color $i$ connects the two vertices of the pair. This creates two connected components $G_1, G_2$, to which (P) applies. The inverse operation does not change $\delta(G)$ (one looses $D$ faces and one connected component), and if there is a non-melonic bubble $B_{\hat{k}}$ in $G_1$ or $G_2$, it gives rise to a non-melonic bubble in $G$. The property (P) thus applies to $G$. In the last case $iii)$, one may cut the two edges of color $i$ and reconnect them as done for case $ii)$. The number of faces then increases by $D-2$ if the vertices of the pair belong to two different faces of color $ij$, and by $D-1$ or $D$ otherwise, in which cases one finds $\delta(G)\ge D-1$ directly. On the other hand, if the vertices of the pair belong to two different faces of color $ij$, two such faces of color $ij$ would still be in a bubble $B_{\hat k}$ with $k\neq i,j$, so that it cannot be melonic. This proves the property (P) and concludes the proof.

\subsection{4-point function: LO, NLO, and so on}

For generic $q$, the four external legs of a 4-point graph have either all the same color, or two legs have one color while the other two have another color. Cutting two edges $e$ and $e'$ in a vacuum graph $G$ gives a 4-point graph $G_{e,e'}$. The other way around there is always a way to glue the external legs of a 4-point graph two by two so as to obtain a vacuum graph. Therefore all 4-point graphs can be realized (non-uniquely) as a graph of the type $G_{e,e'}$.

We denote $\eta(G_{e,e'})$ the number of faces which are broken by cutting $e$ and $e'$, so that the graph $G_{e, e'}$ contributes at order
\begin{equation}
\chi(G_{e,e'}) = \chi(G) - \eta(G_{e, e'}).
\end{equation}
In the SYK model we found $\eta(G_{e, e'})\in\{1, 2\}$ since it only probes the faces with colors $0i$. Here however, $\eta$ can have more values.

\paragraph{Number of broken faces with two external colors.} We first restrict attention to the 4-point function with two colors in its external legs. Cutting first $e$ of color $i\in\{1, \dotsc, q\}$ in $G$ opens up $q-1$ faces (those of colors $(ik)$). Cutting $e'$ of color $j\neq i$ then breaks the $q-2$ faces of color $(jk)$ with $k\neq i$ and possibly another face of color $(ij)$. Therefore
\begin{equation} \label{EtaTwoColors}
\eta(G_{e, e'}) = 2q-3+\epsilon_{e,e'} \qquad \text{with $\epsilon_{e,e'}=0,1$},
\end{equation}
and $\chi(G_{e,e'})= \chi(G) - 2q + 3 - \epsilon_{e,e'} = -q + 2 - \delta(G) - \epsilon_{e,e'}$.

\paragraph{Number of broken faces with one external color.} When all external legs have the same color, there can be more than a single face running along both $e$ and $e'$, which can make $\eta$ less than in the case of two external colors. We can write $\eta(G_{e,e'}) = 2(q-1) - \phi_{e,e'}$ where $\phi_{e,e'}$ is the number of faces shared by $e$ and $e'$.  In the SYK model, we could always restrict attention to cases where $\eta(G_{e,e'})=2$ and disregard the cases where it is 1. Here in a similar way, we can restrict attention to the cases where it is maximal in the following sense.

Given a 4-point graph $G_4$ with the same color on all external lines, we have three ways to glue them two by two and obtain a vacuum graph so that $G_4$ can be obtained from different vacuum graphs $G$ with edges $e,e'$ cut. So we want a prescription to identify a graph $G$ such that $G_4 = G_{e,e'}$ in a canonical way. The prescription is to choose one of the three ways to connect the external lines which maximizes the total number of faces created, i.e. maximizes $\eta$. This gives a vacuum graph $G$ with distinguished edges $e,e'$ and only then we say that the initial 4-point graph is $G_4 = G_{e,e'}$.

Combinatorial considerations\footnote{The proof consists in counting the number of broken faces going from one external line, say the leg $a$, to the others labeled $b, c, d$. There are respectively $\epsilon_b, \epsilon_c, \epsilon_d$ broken faces going from the leg $a$ to the legs $b, c, d$. The total number of broken faces going along the external line $a$ is $\epsilon_b + \epsilon_c + \epsilon_d =q-1$. The three ways to reconnect the external lines two by two can create $f_b = 2\epsilon_b + \epsilon_c + \epsilon_d$ faces, or $f_c = \epsilon_b + 2\epsilon_c + \epsilon_d$, or $f_d = \epsilon_b + \epsilon_c + 2\epsilon_d$ faces. Then $\eta = \max (f_b, f_c, f_d)$ with the constraint $f_b + f_c + f_d = 4(\epsilon_b + \epsilon_c + \epsilon_d) = 4(q-1)$.} show that
\begin{equation} \label{EtaOneColor}
\eta(G_{e,e'}) \geq \frac{4}{3}(q-1)
\end{equation}
in general and 
\begin{equation} \label{EtaOneColorBipartite}
\eta(G_{e,e'}) \geq \frac{3}{2}(q-1)
\end{equation}
for bipartite graphs. This is equivalent to saying that the minimal number of faces shared by $e$ and $e'$ satisfies $\phi_{e,e'} \leq 2(q-1)/3$ in general and $\phi_{e,e'} \leq (q-1)/2$ for bipartite graphs. The reason is that when those inequalities are not satisfied, then it implies that the same 4-point graph can in fact be obtained from another $G_{e_1,e_2}$ with $\eta(G_{e_1,e_2}) > \eta(G_{e,e'})$.

\paragraph{Broken and unbroken chains.} At LO, the cases of one and two colors on the external legs are similar. This is due to a property of melonic graphs (those satisfying $\delta(G) = 0$). In a vacuum melonic graph, two edges $e$ and $e'$ may belong to 1 or no common face. This is obviously true for any (not necessarily melonic) graph if $e$ and $e'$ have different colors. If they have the same color $i$, one might add edges of color 0 on the canonical pairs and contract them as we did in (\ref{Contraction}) for the SYK model. One obtains a graph with no multicolored cycles if $G$ is melonic. The only pairs of vertices which may belong to more than one common face are on the same monocolored cycle, and opening them would therefore separate the graph into two connected components, which is excluded.

This implies that equation \eqref{EtaTwoColors} is true both when there are two colors on the external legs and at LO when there is a single color on the external legs. Setting $\delta(G)=0$, we get two types of contributions depending on the value of $\epsilon_{e,e'}=0,1$. They were described in \cite{Gurau, GS} as unbroken (U) and broken (B) chains,
\begin{equation}
\label{LO_NLO4p}
\begin{array}{c} \includegraphics[scale=.5]{Uchain.pdf} \end{array} \sim N^{2-q} \hspace{3cm} \begin{array}{c} \includegraphics[scale=.5]{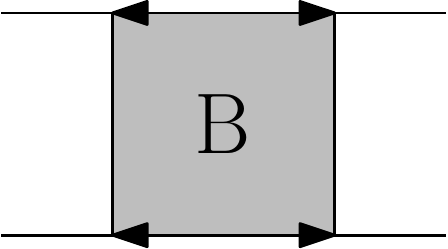} \end{array} \sim N^{1-q}.
\end{equation}
Unbroken chains were defined in \eqref{UnbrokenDef}. A chain is broken if it is not unbroken, or equivalently no faces propagate from one end of the chain to the other.

To prove that there are no other contributions with the same exponents of $N$, $\chi(G_{e,e'}) \geq 1-q$, we simply apply the bounds previously established on the number of broken faces. Assume that $G$ is not melonic. It has exponent $\chi(G) \leq 2$ or even $\chi(G) \leq 1$ for non-bipartite graphs (see Section \ref{sec:2PtGW}). We distinguish the following cases.
\begin{itemize}
\item $\chi(G) \leq 1$ and one external color: then the bound \eqref{EtaOneColor} leads to
\begin{equation}
\chi(G_{e,e'}) \leq 1-q + \frac{1}{3}(4 - q).
\end{equation}
This shows that for $q>4$ all contributions coming from cutting two edges of the same color in a vacuum graph $G$ with $\chi(G)\leq 1$ can be disregarded at orders up to $\chi(G_{e,e'}) \geq 1-q$ (either they contribute to $\chi(G_{e,e'}) <1-q$ or do not maximize $\eta_{e,e'}$ and are thus obtained by cutting edges in different graphs).

However at $q=4$, contributions coming from cutting two edges of the same color in a vacuum graph $G$ with $\chi(G) = 1$ can be expected at order $\chi(G_{e,e'}) = 1-q = -3$. We will in fact exhibit such a contribution in \eqref{GW_NLO4Pt_3D_3}.

\item $\chi(G) = 2$ and one external color: then the graphs are bipartite and the bound \eqref{EtaOneColorBipartite} leads to
\begin{equation}
\chi(G_{e,e'}) \leq 1-q + \frac{1}{2} (5 - q).
\end{equation}
The conclusion is similar to the previous case. For $q>4$, all contributions coming from cutting two edges of the same color in a vacuum graph $G$ with $\chi(G)= 2$ can be disregarded at orders up to $\chi(G_{e,e'}) \geq 1-q$ (either they contribute to $\chi(G_{e,e'}) <1-q$ or do not maximize $\eta_{e,e'}$ and are thus obtained by cutting edges in different graphs).

However at $q=4$, contributions coming from cutting two edges of the same color in a vacuum graph $G$ with $\chi(G) = 2$ can be expected at order $\chi(G_{e,e'}) = 1-q = -3$. We will give examples below of 4-point graphs with $\chi(G_{e,e'}) = 5 - 2q$ which is the same as $1-q$ at $q=4$.

\item $\chi(G) \leq 2$ and two external colors: then the bound \eqref{EtaTwoColors} leads to
\begin{equation}
\chi(G_{e,e'}) \leq 5-2q = 1-q + (4-q).
\end{equation}
The conclusion is similar to the previous two cases. For $q>4$, all contributions coming from cutting two edges of different colors in a vacuum graph $G$ with $\chi(G)\leq 2$ can be disregarded at orders up to $\chi(G_{e,e'}) \geq 1-q$ (they contribute to $\chi(G_{e,e'}) <1-q$).

However at $q=4$, contributions coming from cutting two edges of different colors in a vacuum graph $G$ with $\chi(G) = 2$ can be expected at order $\chi(G_{e,e'}) = 1-q = -3$. We are going to identify all of them in the next paragraph (since $5-2q=1-q$ at $q=4$).
\end{itemize}

\paragraph{Beyond chains, two external colors.} To go further in the $1/N$ expansion, we restrict attention to the case with two colors on the external legs which is much easier to handle thanks to \eqref{EtaTwoColors}. If $G$ is a vacuum NLO graph, $\chi(G)=2$ and thus $\chi(G_{e,e'}) = 5 - 2q -\epsilon_{e,e'}$. In particular, one can obtain contributions which scale like $\chi(G_{e,e'}) = 5-2q$ by cutting edges such that $\epsilon_{e,e'}=0$. With the same reasoning, it is clear that vacuum graphs of higher orders, i.e. $\chi(G) \leq 1$, contribute only to 4-point graphs of higher order, $\chi(G_{e,e'}) \leq 4-2q$.

Let us detail the graphs with $\chi(G_{e,e'}) = 5-2q$: they are obtained by cutting two edges $e,e'$ of different colors $i$ and $j$ in a vacuum NLO graph, with one face of color $(ij)$ going along $e, e'$. One can equivalently cut an edge along a broken face in \eqref{NLO2PtGW} and find
\begin{equation} \label{GW4PtNLO2Pt}
\begin{aligned}
&\begin{array}{c} \includegraphics[scale=.65]{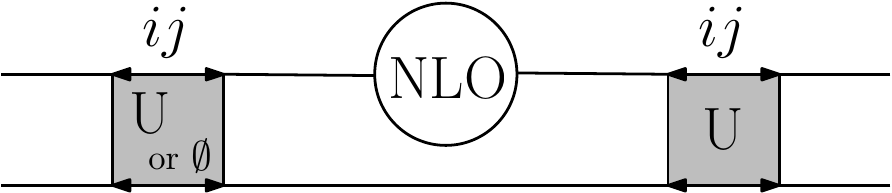} \end{array}
&\begin{array}{c} \includegraphics[scale=.65]{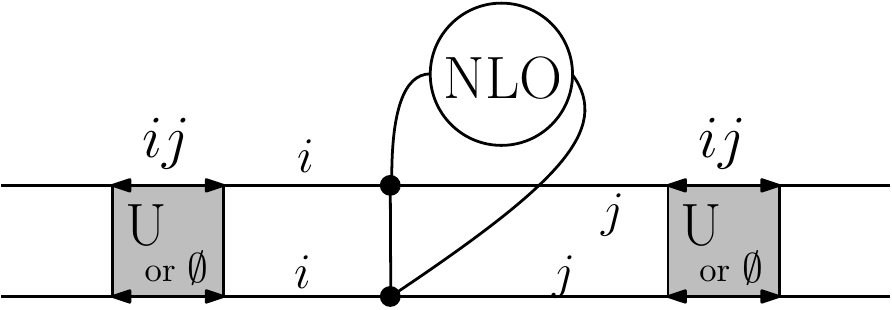} \end{array}
\end{aligned}
\end{equation}
as well as 
\begin{equation} \label{GW4PtGraphs}
\begin{aligned}
&\begin{array}{c} \includegraphics[scale=.65]{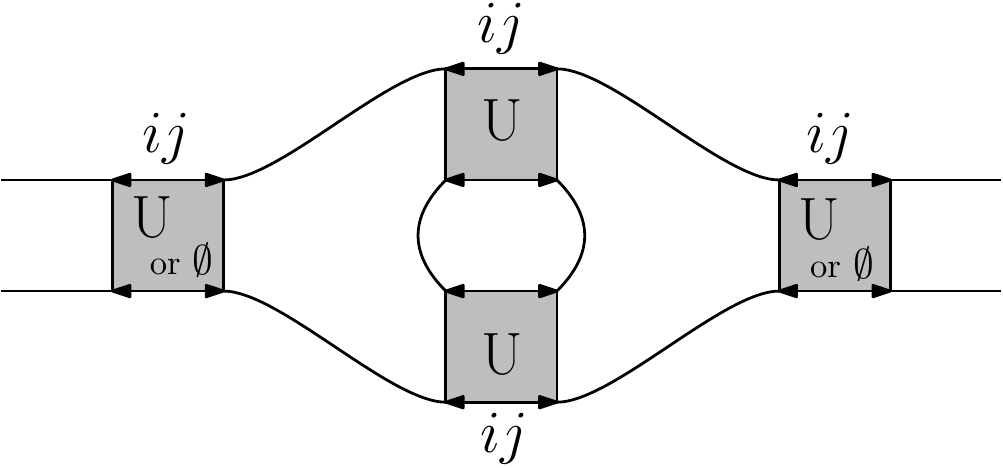} \end{array}
&\begin{array}{c} \includegraphics[scale=.65]{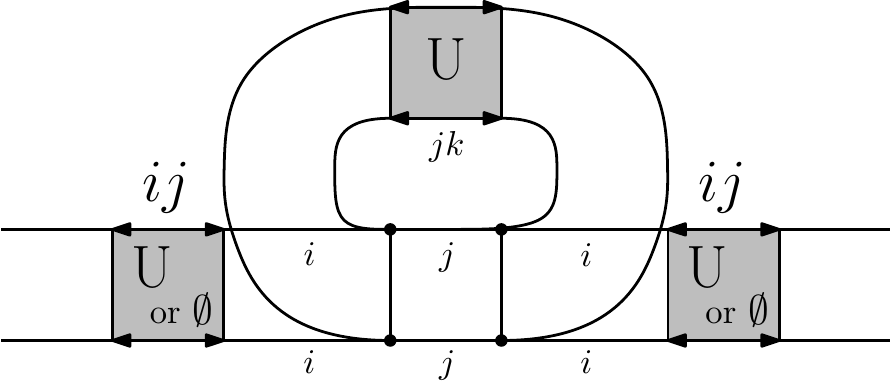} \end{array}\\
&\begin{array}{c} \includegraphics[scale=.65]{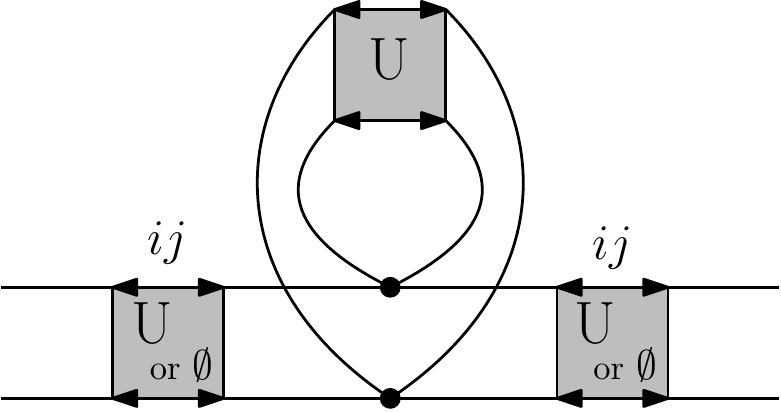} \end{array}
&\begin{array}{c} \includegraphics[scale=.65]{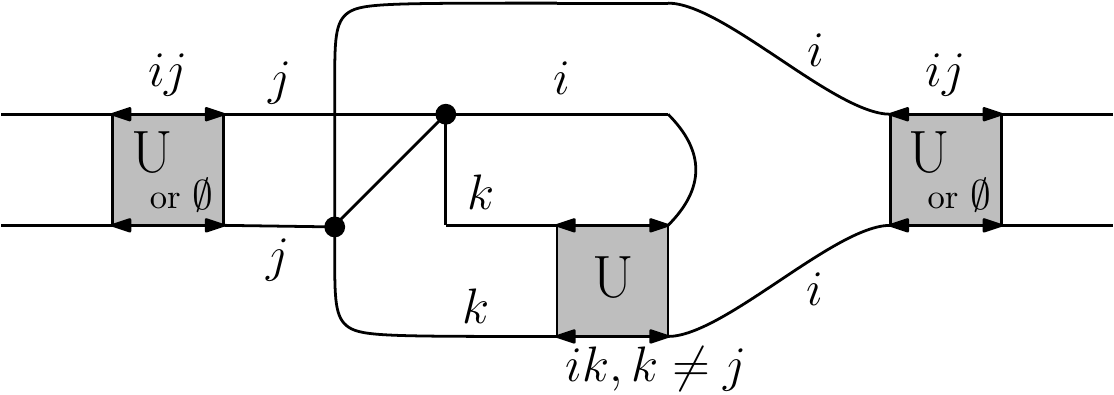} \end{array}\\
&
&\begin{array}{c} \includegraphics[scale=.65]{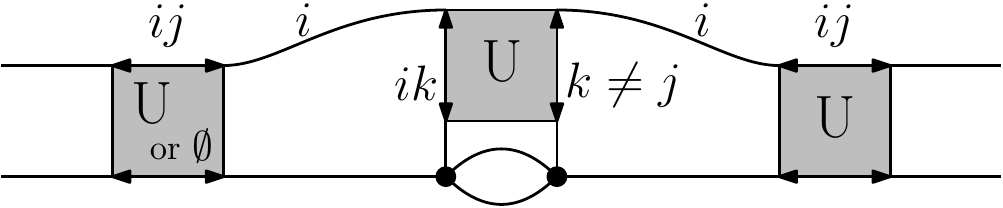} \end{array}
\end{aligned}
\end{equation}
where $i, j$ are the two external colors (it does not matter on which side $i$ is).

Note that for $q=4$, those contributions have the same exponent $\chi(G_{e,e'}) = -3$ as the broken chains since $1-q=5-2q=-3$. They are higher order contributions for $q>4$.

\paragraph{A single external color.} The case with the same color on all four external legs is more difficult as we have not been able to identify the contributions following directly \eqref{LO_NLO4p} in the $1/N$ expansion. We have found a family of graphs with $\chi(G_{e,e'}) = 6-2q$, for any $q>4$ (so it first appears at $q=6$ in the Gurau-Witten model),
\begin{equation}
\label{NNLO4pt}
\begin{array}{c} \includegraphics[scale=.75]{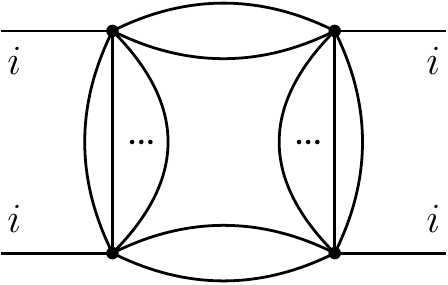} \end{array} 
\end{equation}
for which there are two faces running along both $e$ and $e'$. The graphs of \eqref{GW4PtNLO2Pt} and \eqref{GW4PtGraphs}, amended to have the same colors on the external lines, contribute at $\chi(G_{e,e'}) = 5-2q$. Notice that at $q=4$, they actually contribute with the same exponent of $N$ as the broken chains \eqref{LO_NLO4p} since then $1-q = 5 - 2q = -3$.

Moreover, still at $q=4$, there are NNLO vacuum graphs which, after cutting two edges, also contribute to the 4-point function at the same order $\chi(G_{e,e'})=-3$. An example is the following
\begin{equation}
\label{GW_NLO4Pt_3D_3}
\begin{array}{c} \includegraphics[scale=.75]{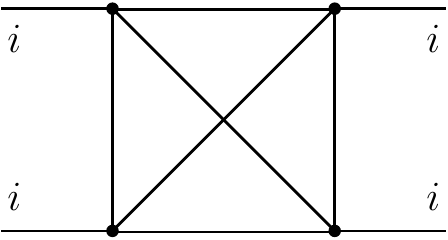} \end{array} 
\end{equation}

However, we have not proved that there are no graphs with $\chi(G_{e,e'})$ between $1-q$ and $6-2q$ for $q>4$ and that there are no other graphs contributing at $\chi(G_{e,e'})=-3$ for $q=4$. 
We expect these issues to be particularly difficult, given the already long proof that \eqref{NLO2PtGW} are the NLO 2-point graphs in \cite{GS}.

\paragraph{Four external colors.} There is an exceptional 4-point function at $q=4$, i.e. four colors, with four distinct colors on the external lines, one per leg. A graph contributing to this exceptional 4-point function can be turned into a vacuum graph by adding a vertex and attaching the four external lines to that vertex. Therefore, all graphs with four external colors are obtained by removing a vertex from a vacuum graph. This breaks exactly one face of each color type $(ij)$, for $i< j \in \{1,2,3,4\}$. The $1/N$ expansion of this 4-point function thus follows that of the free energy.

At NLO, one finds the following diagrams
\begin{equation}
\label{NLO_4pt_4col}
\begin{array}{c} \includegraphics[scale=.6]{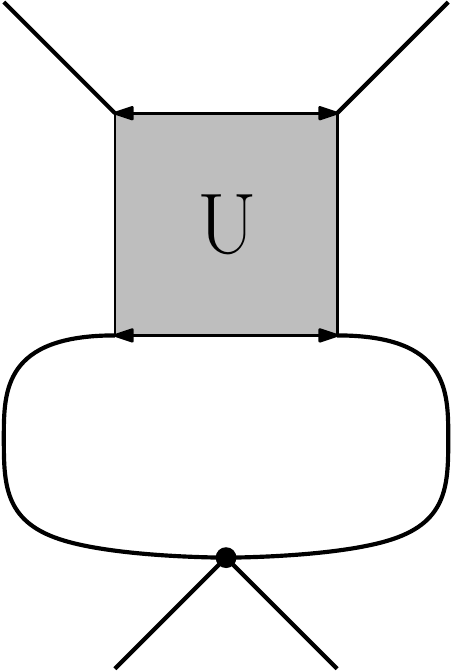} \end{array} \hspace{2cm} \begin{array}{c} \includegraphics[scale=.6]{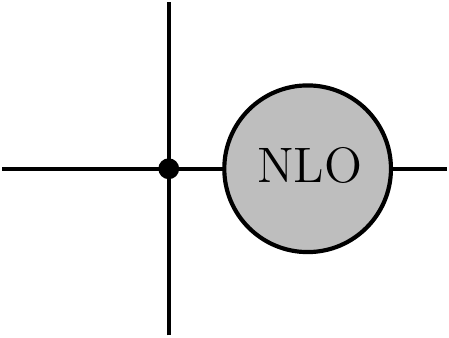} \end{array}
\end{equation}

\section{Summary of results} \label{sec:Summary}

In this section we give a list of the results we have obtained throughout the paper.

\

\noindent{\bf Colored SYK model}

\

\noindent Connected vacuum graphs

\

\begin{tabular}{|c|c|c|c|c|c|}
\hline
{\bf contribution} & ${\bf O(N)}$ &${\bf O(1)}$&${\bf O(1/N)}$&...&${\bf O(1/N^{k-1})}$\\
\hline
{\bf real SYK} & closed melonic& (\ref{Vacuum_NLO_Schemes}) & \begin{tabular}{@{}c@{}}(\ref{VacuumNNLO}),(\ref{SYK_Vacuum_NNLO1}),(\ref{SYK_Vacuum_NNLO2}),(\ref{SYK_Vacuum_NNLO3})\\ and twisted\end{tabular}&...& $\ell_m(G_{/0})=k$\\
\hline
{\bf complex SYK }&"& left of " & (\ref{SYK_Vacuum_NNLO1}),(\ref{SYK_Vacuum_NNLO2}),(\ref{SYK_Vacuum_NNLO3})&...&'', bipartite \\
\hline
\end{tabular}

\

\

\noindent 2-point function

\ 

 \begin{tabular}{|c|c|c|c|}
\hline
{\bf contribution}& ${\bf O(1)}$ &${\bf O(1/N)}$&${\bf O(1/N^2)}$\\
\hline
{\bf real SYK}  & melonic Fig.\ref{melonSYK} & (\ref{NLO2Pt1}) and (\ref{NLO2Pt2}) & \begin{tabular}{@{}c@{}}open an edge in (\ref{SYK_Vacuum_NNLO1}),(\ref{SYK_Vacuum_NNLO2}),(\ref{SYK_Vacuum_NNLO3})\\ and twisted\end{tabular}\\
\hline
{\bf complex SYK } & '' & left of " & open an edge in (\ref{SYK_Vacuum_NNLO1}),(\ref{SYK_Vacuum_NNLO2}),(\ref{SYK_Vacuum_NNLO3}) \\
\hline
\end{tabular}

\

\noindent 4-point function

\

\begin{tabular}{|c|c|c|c|}
\hline
{\bf contribution} & ${\bf O(1/N)}$ &${\bf O(1/N^2)}$&${\bf O(1/N^3)}$\\
\hline
{\bf real SYK}  & (\ref{SYKChainVertex}) & (\ref{NLO4Pt1A}), (\ref{NLO4Pt1B}), (\ref{NLO4Pt1C}), (\ref{NLO4Pt1D}), (\ref{NLO4Pt1E}) & \begin{tabular}{@{}c@{}}open two edges in (\ref{SYK_Vacuum_NNLO1}),(\ref{SYK_Vacuum_NNLO2}),(\ref{SYK_Vacuum_NNLO3})\\ and twisted\end{tabular} \footnotemark[2]\\
\hline
{\bf complex SYK }& '' & non-twisted in " & open two edges in (\ref{SYK_Vacuum_NNLO1}),(\ref{SYK_Vacuum_NNLO2}),(\ref{SYK_Vacuum_NNLO3}) \footnotemark[2]\\
\hline
\end{tabular}
\footnotetext[2]{\label{note1}Cut edges $e$ and $e'$ such that $\eta(G_{e,e'})=2$.}

\

\hspace{0.5cm}

\noindent{\bf Gurau-Witten model}

\

\noindent Connected vacuum graphs

\

 \begin{tabular}{|c|c|c|c|c|c|}
\hline
{\bf contribution} & ${\bf O(N^{q-1})}$ &${\bf O(N^2)}$&${\bf O(N)}$&${\bf O(1)}$\\
\hline
{\bf \begin{tabular}{@{}c@{}}real\\ Gurau-Witten \end{tabular}} & \begin{tabular}{@{}c@{}}closed\\melonic\end{tabular}& \begin{tabular}{@{}c@{}}left of (\ref{Vacuum_NLO_Schemes}), without\\ color 0, and non-\\ separable ``U" chain\end{tabular}&\begin{tabular}{@{}c@{}}right of (\ref{Vacuum_NLO_Schemes}), without\\ color 0, and non-\\ separable ``U" chain\end{tabular}&\begin{tabular}{@{}c@{}}(\ref{Vacuum_NLO_Schemes}) with broken ``B"\\ chains instead of ``U" \end{tabular}\\
\hline
{\bf \begin{tabular}{@{}c@{}}complex\\ Gurau-Witten \end{tabular} }&"& " & $\emptyset$ &\begin{tabular}{@{}c@{}}left of " \end{tabular} \\
\hline
\end{tabular}

\

\

For $D>4$, the previous contributions are the LO, NLO, NNLO and NNNLO. We do not prove that those NNLO and NNNLO are the only ones. For $D=3,4$, there are other obvious contributions as well to the NNLO (e.g. non-twisted (24), (25), (26) without the color 0 edges and such that non-separable chains are of the unbroken type) and to the NNNLO (e.g. (24), (25), (26) without the color 0 edges and such that non-separable chains are of the unbroken type, one of them containing a twist).

\


\noindent 2-point function

\

 \begin{tabular}{|c|c|c|c|c|c|}
\hline
{\bf contribution} & ${\bf O(1)}$ &${\bf O(1/N^{q-3})}$&${\bf O(1/ {N^{q-2})}}$&${\bf O(1/N^{q-1})}$\\
\hline
{\bf \begin{tabular}{@{}c@{}}real\\ Gurau-Witten \end{tabular}} & melonic Fig.\ref{melonGW} & (\ref{NLO2PtGW})  &(\ref{NNLO2PtGW})&\begin{tabular}{@{}c@{}}(\ref{NLO2PtGW}) and (\ref{NNLO2PtGW}), with broken\\ ``B" chains instead of ``U" \end{tabular}\\
\hline
{\bf \begin{tabular}{@{}c@{}}complex\\ Gurau-Witten \end{tabular}}&"& " & $\emptyset$ &\begin{tabular}{@{}c@{}}(\ref{NLO2PtGW}), with broken\\ ``B" chains instead of ``U" \end{tabular} \\
\hline
\end{tabular}

\

\

\noindent 4-point function $q\ge 6$ - two colors

\

 \begin{tabular}{|c|c|c|c|}
\hline
{\bf contribution} & ${\bf O(1 /{N^{q-2}})}$ &${\bf O(1 /{N^{q-1}})}$&${\bf O(1 /{N^{2q-5}})}$\\
\hline
{\bf real Gurau-Witten} & \begin{tabular}{@{}c@{}}unbroken chain\\left of (\ref{LO_NLO4p})\end{tabular}& \begin{tabular}{@{}c@{}}broken chain \\right of (\ref{LO_NLO4p})\end{tabular} &(\ref{GW4PtNLO2Pt}) and (\ref{GW4PtGraphs}) \\
\hline
{\bf complex Gurau-Witten}&"& " & "  \\
\hline
\end{tabular}

\

\

\noindent 4-point function $q=4$ - two colors

\

 \begin{tabular}{|c|c|c|c|c|c|}
\hline
{\bf contribution} & ${\bf O(1 /{N^{2}})}$ &${\bf O(1/ {N^3})}$\\
\hline
{\bf real Gurau-Witten} & \begin{tabular}{@{}c@{}}unbroken chain\\left of (\ref{LO_NLO4p})\end{tabular}& \begin{tabular}{@{}c@{}}broken chain, right of (\ref{LO_NLO4p})\\ (\ref{GW4PtNLO2Pt}) and (\ref{GW4PtGraphs}) \end{tabular}  \\
\hline
{\bf complex Gurau-Witten}&"& "   \\
\hline
\end{tabular}

\

\

\noindent 4-point function $q\ge6$ - one color

\

 \begin{tabular}{|c|c|c|c|}
\hline
{\bf contribution} & ${\bf O(1 /{N^{q-2}})}$ &${\bf O(1 /{N^{q-1}})}$&${\bf O(1 /{N^{2q-6}})}$\\
\hline
{\bf real Gurau-Witten} & \begin{tabular}{@{}c@{}}unbroken chain\\left of (\ref{LO_NLO4p})\end{tabular}& \begin{tabular}{@{}c@{}}broken chain \\right of (\ref{LO_NLO4p})\end{tabular} &(\ref{NNLO4pt}) \\
\hline
{\bf complex Gurau-Witten}&"& " & "  \\
\hline
\end{tabular}

\

\

\noindent 4-point function $q=4$ - one color

\

 \begin{tabular}{|c|c|c|c|c|c|}
\hline
{\bf contribution} & ${\bf O(1 /{N^{2}})}$ &${\bf O(1/ {N^3})}$\\
\hline
{\bf real Gurau-Witten} & \begin{tabular}{@{}c@{}}unbroken chain\\left of (\ref{LO_NLO4p})\end{tabular}& \begin{tabular}{@{}c@{}}broken chain, right of (\ref{LO_NLO4p})\\ (\ref{GW4PtNLO2Pt}), (\ref{GW4PtGraphs}),  (\ref{GW_NLO4Pt_3D_3}) \end{tabular}  \\
\hline
{\bf complex Gurau-Witten}&"& " \quad except (\ref{GW_NLO4Pt_3D_3})  \\
\hline
\end{tabular}

\

\

\noindent 4-point function $q=4$ - four colors

\

 \begin{tabular}{|c|c|c|c|c|c|}
\hline
{\bf contribution} & ${\bf O(1 /{N^{3/2}})}$ &${\bf O(1/ {N^{5/2}})}$\\
\hline
{\bf real Gurau-Witten} & \begin{tabular}{@{}c@{}} A single vertex with four external \\ legs dressed with propagators \end{tabular}& (\ref{NLO_4pt_4col})\\
\hline
{\bf complex Gurau-Witten}&"& "   \\
\hline
\end{tabular}

\section{Concluding remarks} \label{sec:Conclusion}

In this paper we have proposed in Section \ref{sec:GeneralizedSYK} a colored version of the SYK model, which is a particular case of the Gross and Rosenhaus generalization of the SYK model \cite{Gross}, and is a real version of the one by Gurau in \cite{gurau-ultim}. We have analyzed in Section \ref{sec:SYK} the diagrammatics of the two- and four-point functions of this model, exhibiting the LO and NLO diagrams in the large $N$ expansion. In fact, we have exposed a method, alternative to the one used by Gurau in \cite{gurau-ultim}, which recasts the diagrammatics in terms of colored cycles in (Eulerian) graphs, see Section \ref{sec:2Pt}. In principle, this method can be carried on to higher orders (the main difficulty being then the larger number of diagrams to account for). In particular, we have applied it to extract the NLO of the 4-point function, going beyond chain diagrams (also called ladders in \cite{MS}) for the first time (although in the context of a colored model).

In Section \ref{sec:GW}, we have performed a similar analysis for the Gurau-Witten model \cite{Witten}. The use of real fermionic fields requires to deal with non-bipartite Feynman graphs in contrast with \cite{Gurau}. We have shown that at LO and NLO for the 2-point function, this model is actually equivalent to the complex one of \cite{Gurau}. We have also presented a similar analysis for the 4-point function, going beyond the contributions of chains (introduced in \cite{GS}) for the first time.

Remarkably, the diagrammatic analysis is much simpler in the colored SYK model than in the Gurau-Witten model, thanks to the method we developed in Section \ref{sec:SYK}. Furthermore, the analysis of the 4-point function in the Gurau-Witten model depends on the number of distinct colors on the external lines and it becomes more involved the fewer colors are present.

As mentioned in the introduction, we know the two models are different in general but the interest in the Gurau-Witten model is that its LO coincide with the SYK model. At NLO (for the free energy or the 2-point function), the two models start to differ although the diagrams remain quite similar.

In fact, we have found that diagrams which contribute to the same order in the SYK model can contribute to different orders in the Gurau-Witten model. For example, in the case of the 4-point function for the Gurau-Witten model, we have a distinction between broken and unbroken chains, phenomenon which is not existing in the case of the colored SYK model. The Gurau-Witten thus sort of lifts a degeneracy of Feynman graphs of the SYK (at least at low orders). This is obviously due to the fact that both models have different exponents of $N$, taking into account different faces. This phenomenon is likely to be more and more important when one studies further and further orders (NNLO, NNNLO and so on) in the diagrammatics of the large $N$ expansions of the two models. 

A first perspective for future work appears to us to be the computation of the corresponding Feynman amplitudes of the diagrams we have shown in this paper. It would be interesting to see if the lifting of degeneracy in the Gurau-Witten model mentioned above has some physical effect on the sum of diagrams at NLO, for both types of models analysed here.

Moreover, as already mentioned in the introduction, several other SYK-like tensor models exist in the literature (the Klebanov-Tarnopolsky model \cite{KT}, or the supersymetric model \cite{brown}). It would thus be interesting to apply the diagrammatic techniques we have developed here for the study of these models as well, in order to compare the LO and NLO behavior of all these SYK-like models.

Another perspective of interest as far as the diagrammatics is concerned follows from the following fact: the SYK model is simpler to deal with than the Gurau-Witten model because it only takes into account faces of colors $(0i)$ for $i=1, \dotsc, q$ (compared to $(ij)$ in the Gurau-Witten). There exist tensor models, which are 1-tensor models generalizing the regular 1-matrix models, which have exactly this behavior \cite{uncolored}. An interesting question is whether such models exhibit the same physical features (chaos, conformal invariance) as the SYK model.

\section*{Acknowledgements}
The authors acknowledge Razvan Gurau for several important discussions on the $D=3$ Gurau-Witten model.
AT is partially supported by the grant ANR JCJC ``CombPhysMat2Tens" and by the grant PN 
16 42 01 01/2016. VB is partially supported by the ANR MetACOnc project ANR-15-CE40-0014. The authors also acknowledge the Institute Henri Poincar\'e (IHP) "Combinatorics and Interactions" trimester - part of this work was done at IHP during this trimester.



%
%
%
%
%
	
\end{document}